\documentclass[conference]{IEEEtran}
\usepackage[pdftex]{graphicx}
\usepackage{amsmath}
\usepackage{amsfonts}
\usepackage{eqparbox}
\usepackage{graphicx} 
\usepackage{multicol}
\usepackage{booktabs} 
\usepackage{xmpmulti}
\usepackage{dirtytalk}
\usepackage[T1]{fontenc}
\usepackage{lastpage}
\usepackage{amssymb}

\usepackage{array}

\begin{document}
\title{Lab 4: Design of the Single Balanced Mixer (Using HSMS-281x Diode) and Measurements}
\author{Varun Mannam, \textit{Student Member IEEE}}
\maketitle
\thispagestyle{plain}
\pagestyle{plain}

\begin{abstract}

We designed the Mixer operating at 2.4 GHz for WIFI-band using ADS software and fabricated using Rogers 4350B substrate. We designed the single-balanced mixer using the diode (HSMS-281x model). We measured the isolation across all ports of the mixer. We measured the conversion loss of the mixer. We performed the non-linear measurements, which includes IM3 vs frequency. The performance metrics are compared with the data-sheet values and explained if any discrepancy exists.\newline

\end{abstract}

%%%%%%%%%%%%%%%%%%%%%%%%%%%%%%%%%%%%%%%%%%%%%%%%%%%%%%%%%%%%%%%%%%%%%%%%%%%%%%%%
\section{INTRODUCTION}

The mixer is a critical component of RF transceiver design. The following figure shows the RF receiver block diagram which has the mixer acts as down-converter from RF (high frequency) to IF (low-frequency). 
\begin{figure}[h!]
\centering
\includegraphics[width=8cm]{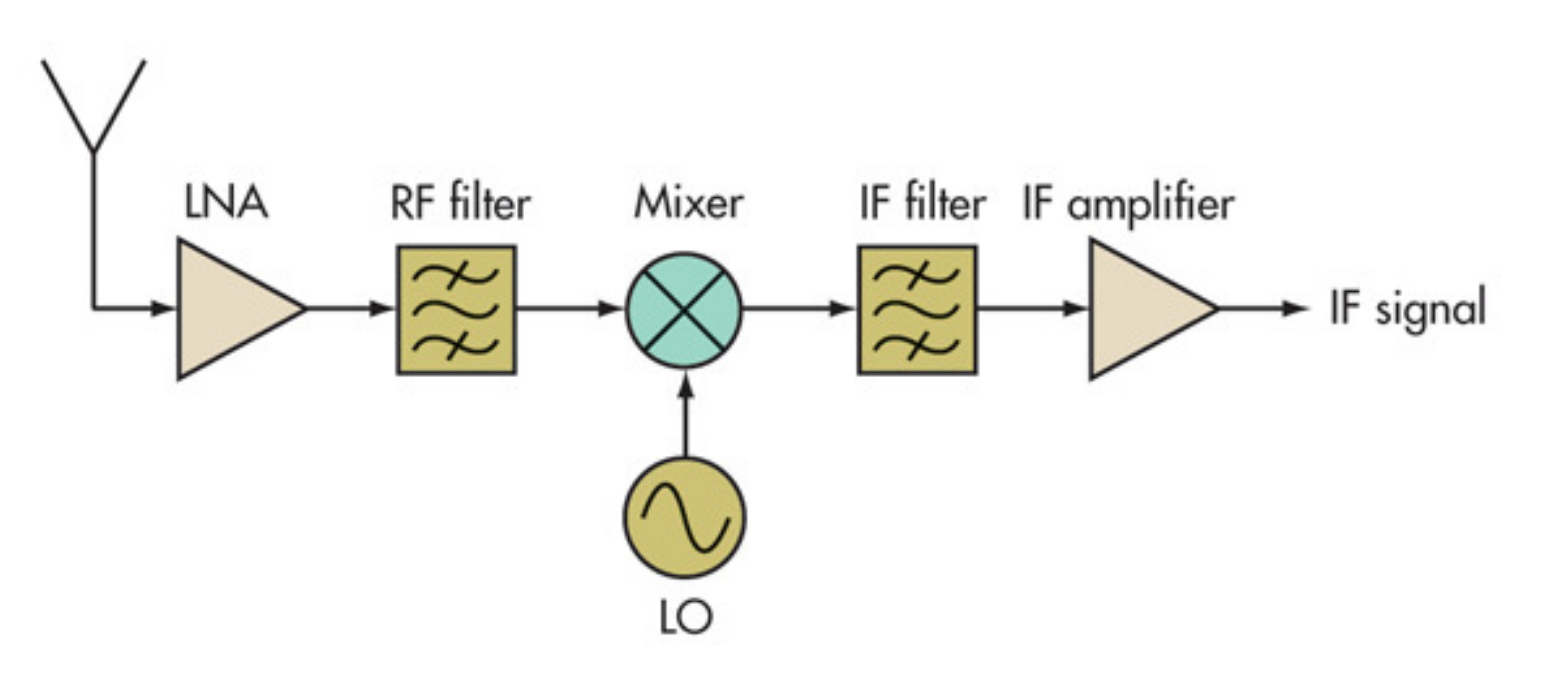}
\caption{RF receiver block diagram}
\end{figure} \\

The mixer is the non-linear device which produces the  non-linear frequencies for up-conversion/down-conversion.The mixer circuit generates the following frequencies: $fIF = n * fLO \pm m * fRF$ (m and n are ALL integers). The Mixer is evolved from single simple diode where RF and LO signal is given to anode and result IF is at cathode port is taken. Now we used balanced structures (like: hybrid) in some combinations. here in-phase and out-of phase signals will be produced using hybrid, where the out of phase signals can be canceled by using some components. The signal is spread across multiple inter-mods in frequency domain. The initial single-ended mixer is designed and added few components to cancel out-of phase inter-mods.The phase-correlation power-diving of 0 and 180 deg are used. In this paper we are designing the \textit{\textbf{single-balanced}} mixer which is the hybrid junction with 2-diode. Here the balance means the measure of cancellation of single-tone inter-mods b/w RF and LO. This mixer provides Isolation b/w RF and LO ports, 50 percent reduction in inter-mods and higher conversion efficiency.\\

The Paper mainly discuss in 2-parts, where as in part-I discuss mainly about the design of Mixer which includes the fabrication and part-II discuss mainly about the measurements on the designed mixer. 
We measured few parameters of the designed Mixer. Those are linear and non-linear measurements. In our lab, professor gave us a HSMS-2841x diode model and attached this to the designed board which is operating at 2.4 GHz. This board has RF ports for inputs (RF and LO) and output (IF). \\

The design of Mixer involves multiple steps mainly, diode IV curves, Large-signal S-parameters of the diode, Rat-race coupler, IF filter and match network for diode to rat-race coupler. Each of the step is explained in the later sections in detail. The Fabrication is done with LPKF machine using Rogers 4350B substrate. The fabrication process is explained separately in the later sections in-detail. \\

From this data sheet of the HSMS-281x, the diode designed is operating at 2.45 GHz using SOT-23 package. The linear model of the diode is taken from Avago Technologies application note 1124. \\

In linear operation, conversion gain is independent of RF signal power. i.e 1dB decrease in RF power results, 1dB decrease in IF power, results same conversion loss, but at high RF power, this effects is not same. The IF output is decreases more than conversion loss of the mixer. This stage is called Compression in mixer. At this stage, the RF power is used as a function of switches along with LO powers. The linear measurements are the conversion loss of the mixer. We measured the isolation of the mixer across all ports. The non-linear measurements are IM3, 1-dB compression point. These measurements are performed using 2-tone test which we did similar to our previous labs.\\

We used a couple of attenuators in our circuit to protect the Mixer and test equipment. The spectrum analyzer handles maximum power of 1watt. We added one more 10dB attenuators with lower power rating. Accounting the cable losses in all measurements are important. We designed the Mixer using HSMS-281x diode and measured the RF performance. The performance metrics are compared with the data-sheet values and explained if any discrepancy exists.\newline

\section{Design of the Mixer}
As mentioned in the Introduction, design of mixer includes the following steps mainly Diode-model, I-V characteristics, Large-signal S-parameters, Rat-race coupler, IF filter design and match network for diode and IF stage. In this section each step of diode design is explained in-detail.\\

\subsection{Diode Model}
In this section, we  created the diode from die model of HSMS-281x data-sheet and package model from Avago-SMT respectively. We created the spice model of the die using ADS software. The spice model parameters are given below.
\begin{table}[h!]
\begin{center}
 \begin{tabular}{|c|c|} 
 \hline
 Parameter (Units) & Value  \\ 
 \hline
 $B_v$ (V) & 25 \\ 
 \hline
 $C_j0$ (pF) & 1.1 \\
 \hline
 $E_g$ (eV) & 0.69 \\
 \hline
 $I_bv$ (A)  & 1e-5 \\
 \hline
 $I_s$ (A) & 4.8e-9 \\
 \hline
 N & 1.08 \\
 \hline
 $R_s$ (Ohm) & 10 \\
 \hline
 M & 0.5 \\
 \hline
 $V_j$ (V) & 0.65 \\
 \hline
 XTI & 2 \\
 \hline
\end{tabular} 
\end{center} 
\caption{SPICE parameters of the diode}
\end{table} \\
The following figure shows the ADS model of the diode.
\begin{figure}[h!]
\centering
\includegraphics[width=8cm]{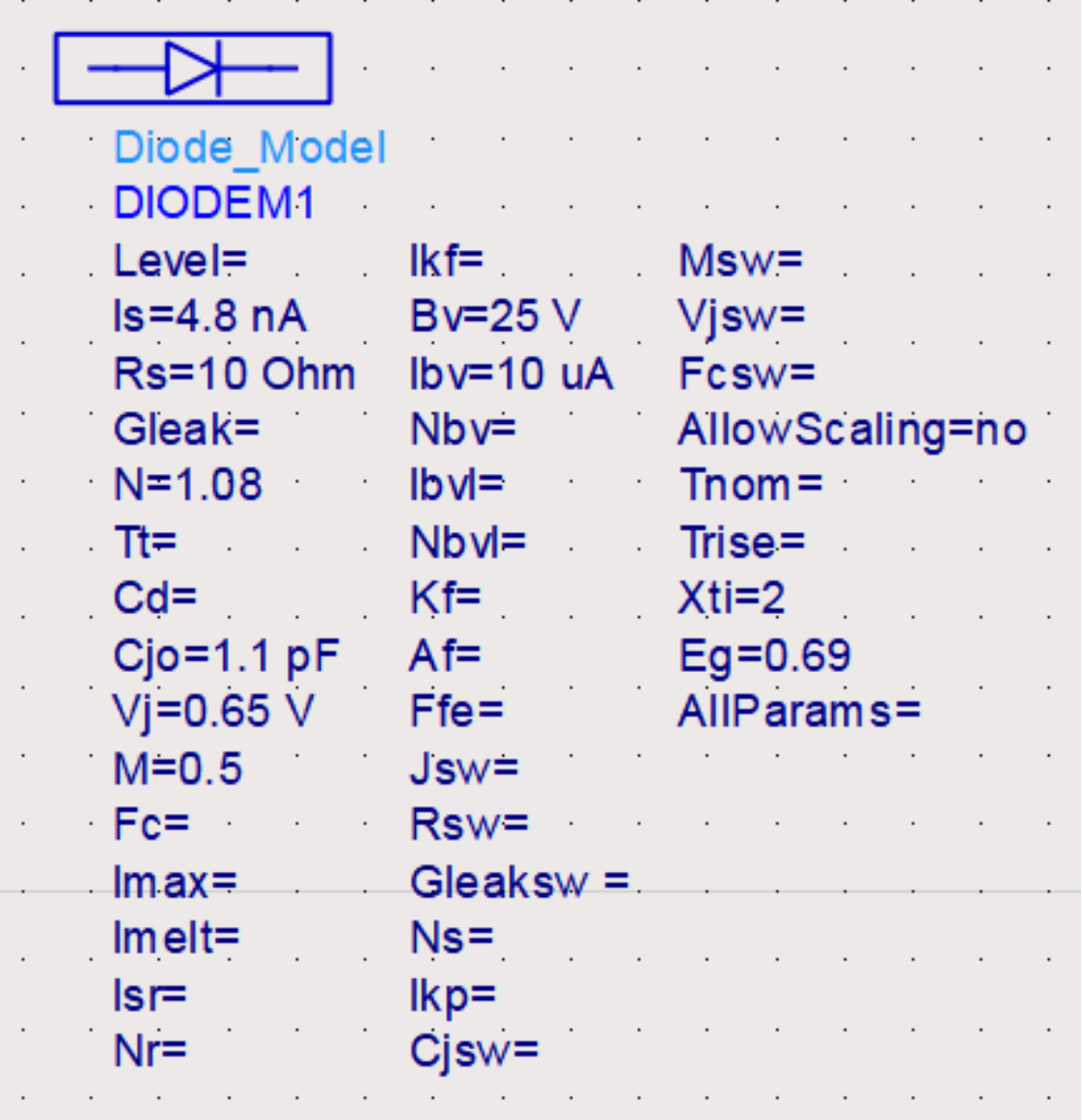}
\caption{HSMS-281x diode model in ADS}
\end{figure} \\

We used the Avago SMT package Model to model the parasitics of the diode which uses the inductor and capacitors. we are using SOT-23 package which includes the two-diodes in the package, out of which only 1-diode is activated at present. The following figure shows the parasitic values of $C_P$,$C_C$ and $C_L$ values. 
\begin{figure}[h!]
\centering
\includegraphics[width=8cm]{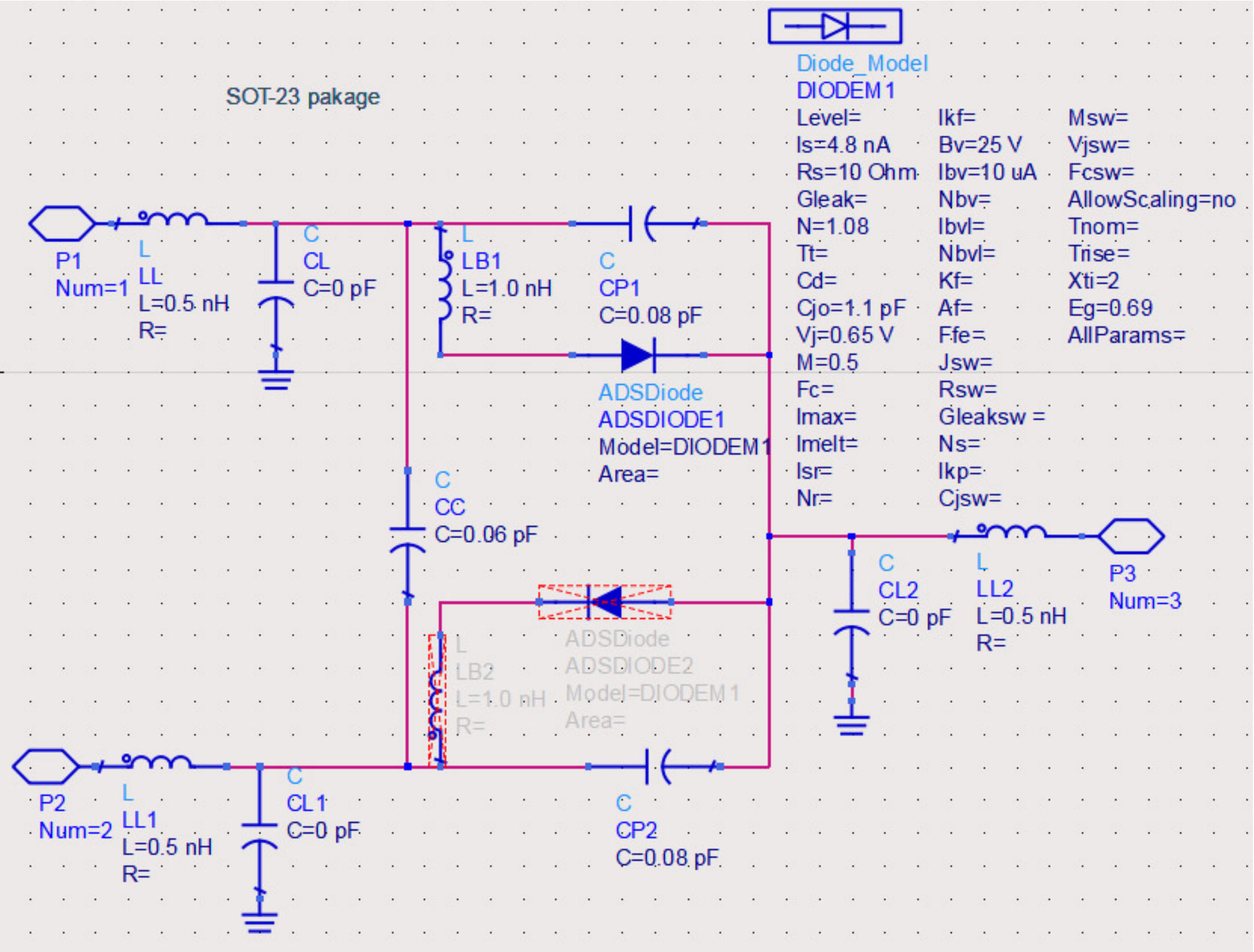}
\caption{HSMS-281x diode package in ADS}
\end{figure} \\

The parasitic values used in the package model are $L_L$ is 0 nH, $C_L$ is 0 pF, $C_P$ is 0.08 pF, Coupling capacitor $C_C$ is 0.06 pF and bond-wire inductance $(L_B)$ is 1.0 nH. \\ 

\subsection{I-V characteristics}
In this section, we performed the DC-simulation of the diode which mainly says the IV characteristics of the diode. The given ADS template have this step which plots the I-V characteristics. The following figure shows the dc-simulation setup of the diode using ADS.
\begin{figure}[h!]
\centering
\includegraphics[width=8cm]{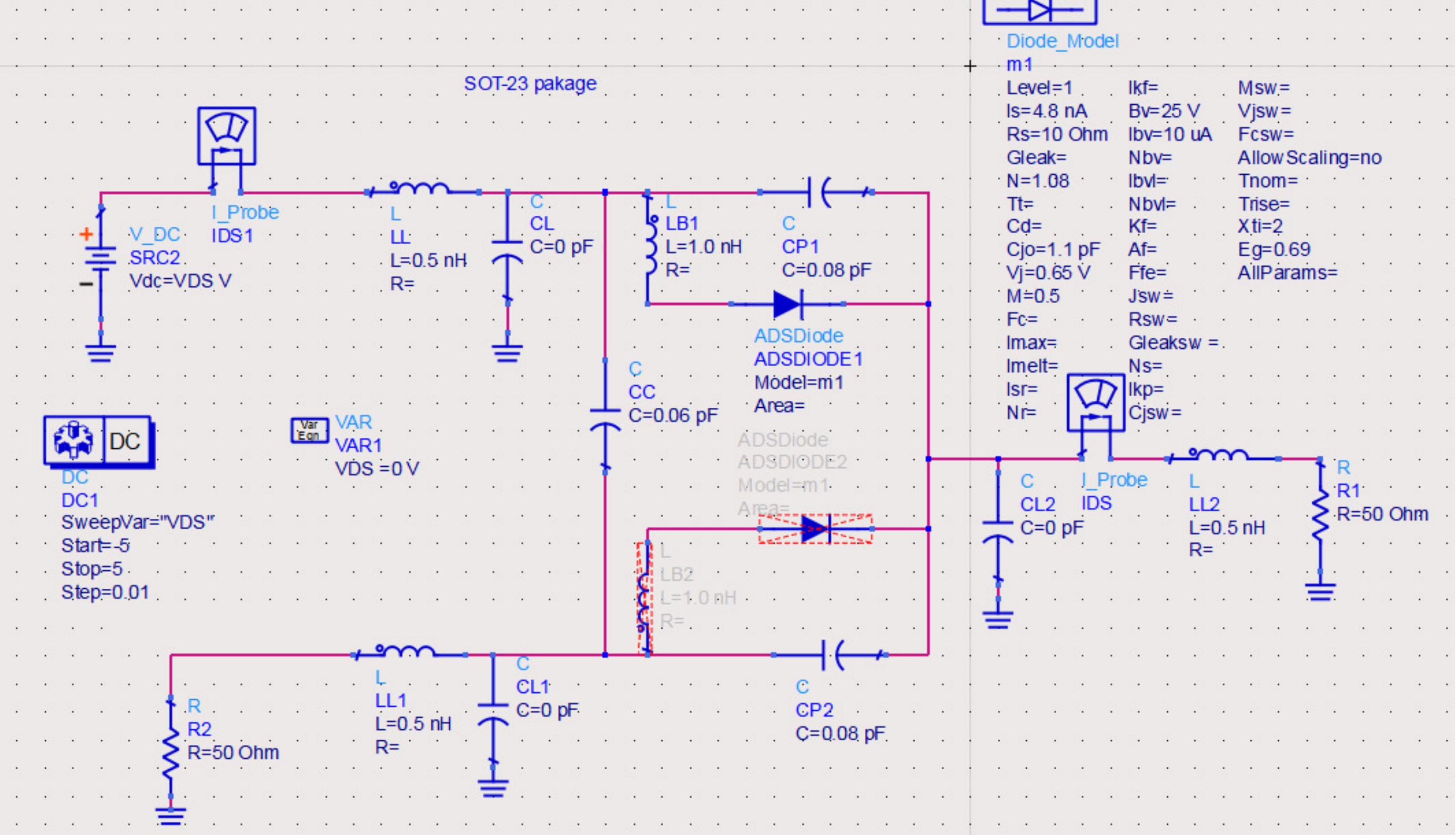}
\caption{DC simulation of diode in ADS}
\end{figure} \\
The following figure shows the IV characteristics of the diode from ADS. The IV curves dives into the 3- regions, first one is the saturation current below $v_D$ of the diode where as the second one the diode current is dominated by input voltage from the effect of junction resistance ($R_j$) and the third one is diode current by linear voltage by the Resistance ($R_s$). These characteristics can be seen in log-log plot of IV curves. 
\begin{figure}[h!]
\centering
\includegraphics[width=9cm]{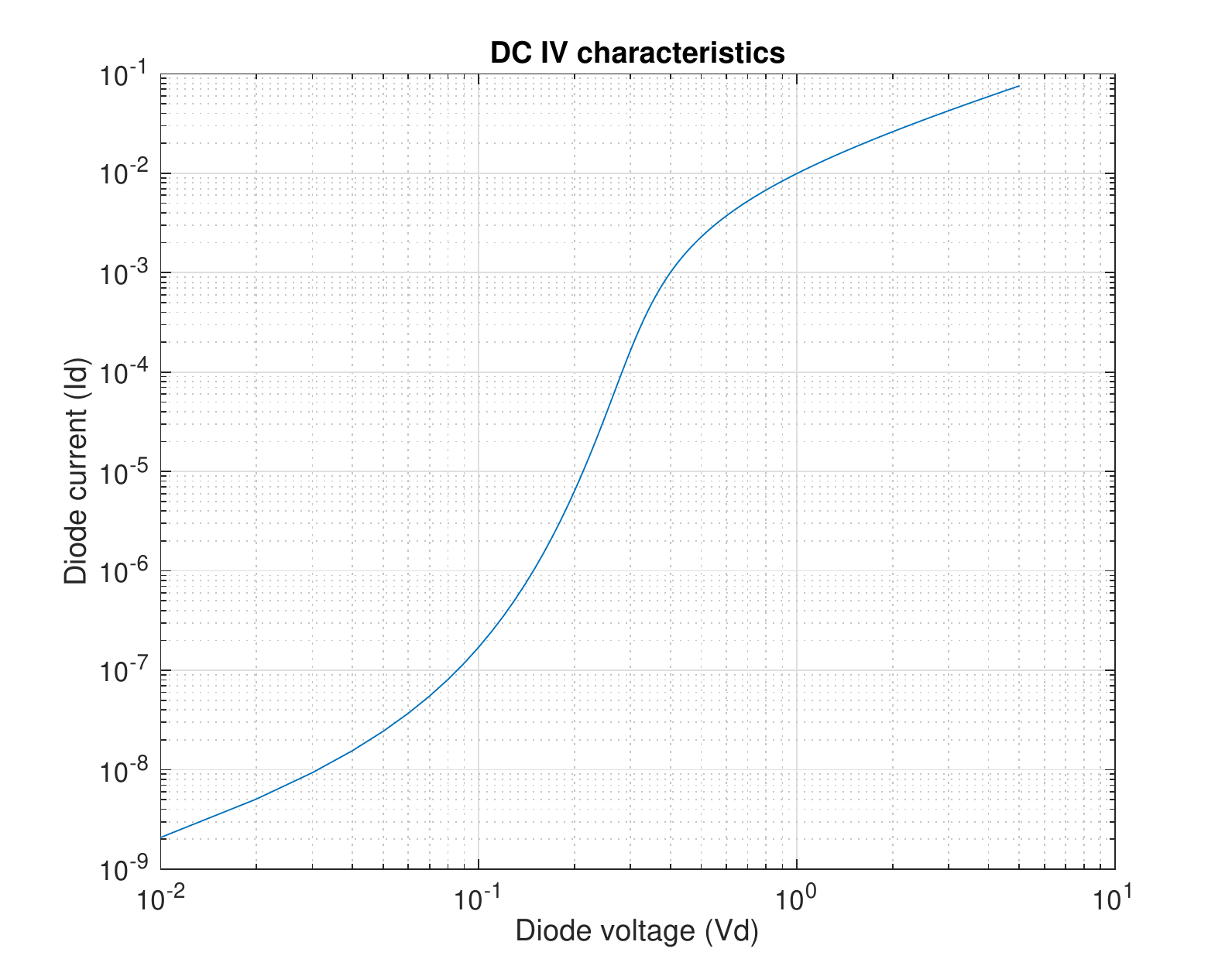}
\caption{I-V curves of the diode in ADS}
\end{figure}
The diode has series resistance ($Rs$) of 10 ohms. \\

The diode equivalent linear model is shown below.Here the $R_s$ is the series resistance, $C_j$ is the junction capacitance and $R_j$ is the junction resistance is given by the following equation.
\begin{equation}
R_j = 8.33*(1e-5)*N*T/(i_b+i_s)
\end{equation}
Here $i_b$ is the bias-current, $i_s$ is the saturation current, N is the ideality factor and T is the temperature.
\begin{figure}[h!]
\centering
\includegraphics[width=5cm]{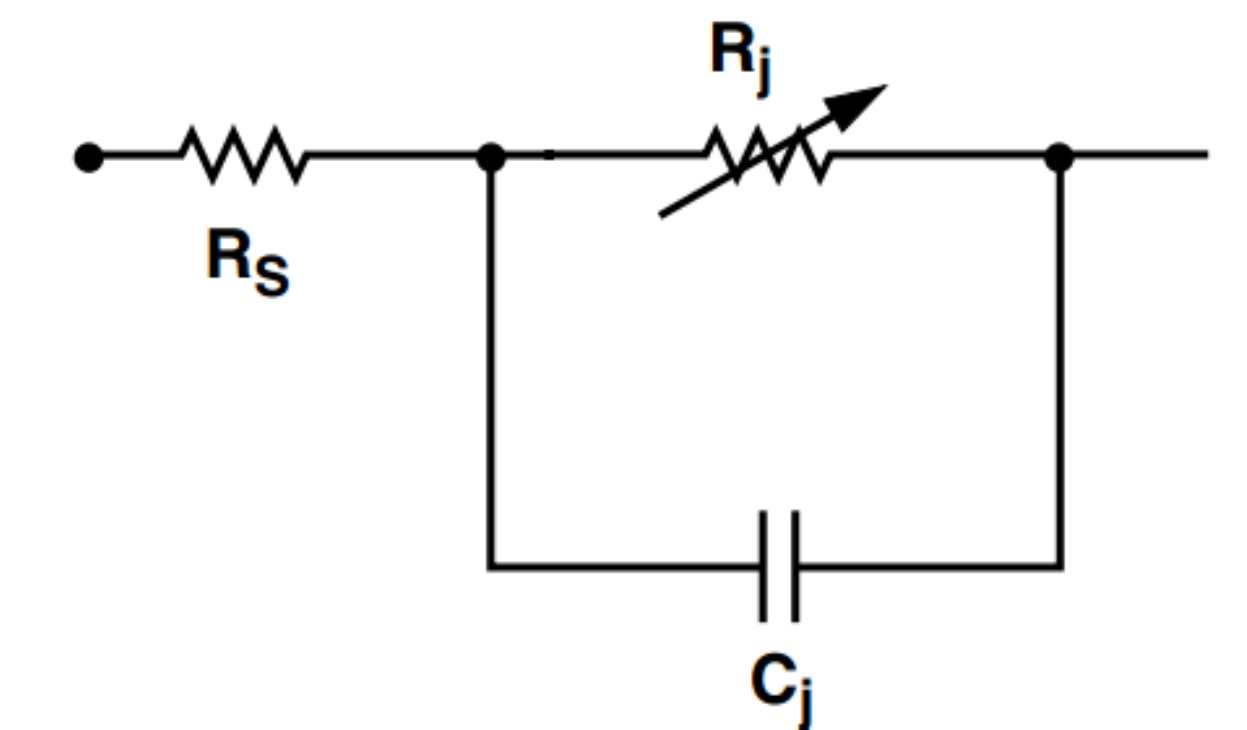}
\caption{Diode model equivalent circuit}
\end{figure} \\

After completing this section we understood the the diode resistance of series and junction values are critical for the operation. The IV curves of the diode are mainly due to these $R_j$ and $R_s$ values.\\

\subsection{Large-signal S-parameters}
In this section, we applied single tone at the diode input and finds the input impedance of the diode at various power levels of the tone. This step is required to know the mismatch of the diode and rat-race coupler section. The following figure shows the schematic of the Large-signal S-parameters (LSSP) of the diode. \\
\begin{figure}[h!]
\includegraphics[width=8cm]{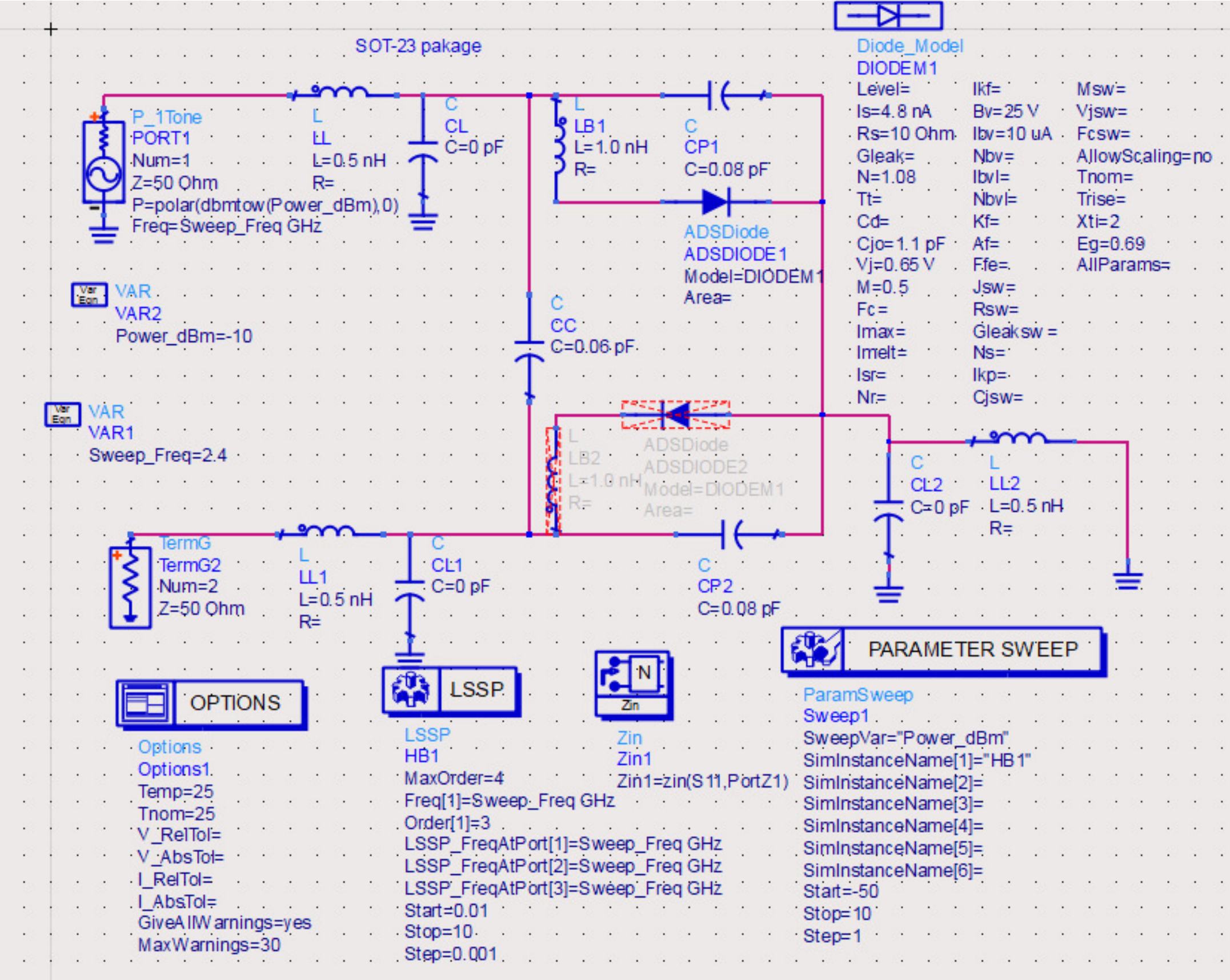}
\caption{LSSP schematic of the diode in ADS}
\end{figure}
From the figure, it is clear that we vary the power level of the input RF tone and compute the $Z_{in}$ value for all power levels. The following figure shows the $Z_{in}$ value for various power levels. Here we select the LO power is 10 dBm, so used $Z_{in}$ corresponding to 10 dBm RF tone power which gives the $Z_{in}$ of 0.573@-110.5 deg at 2.4 GHZ frequency. This $Z_{in}$ is the normalized one with $Z_{0}$ of 50 Ohms. 
\begin{figure}[h!]
\includegraphics[width=8cm]{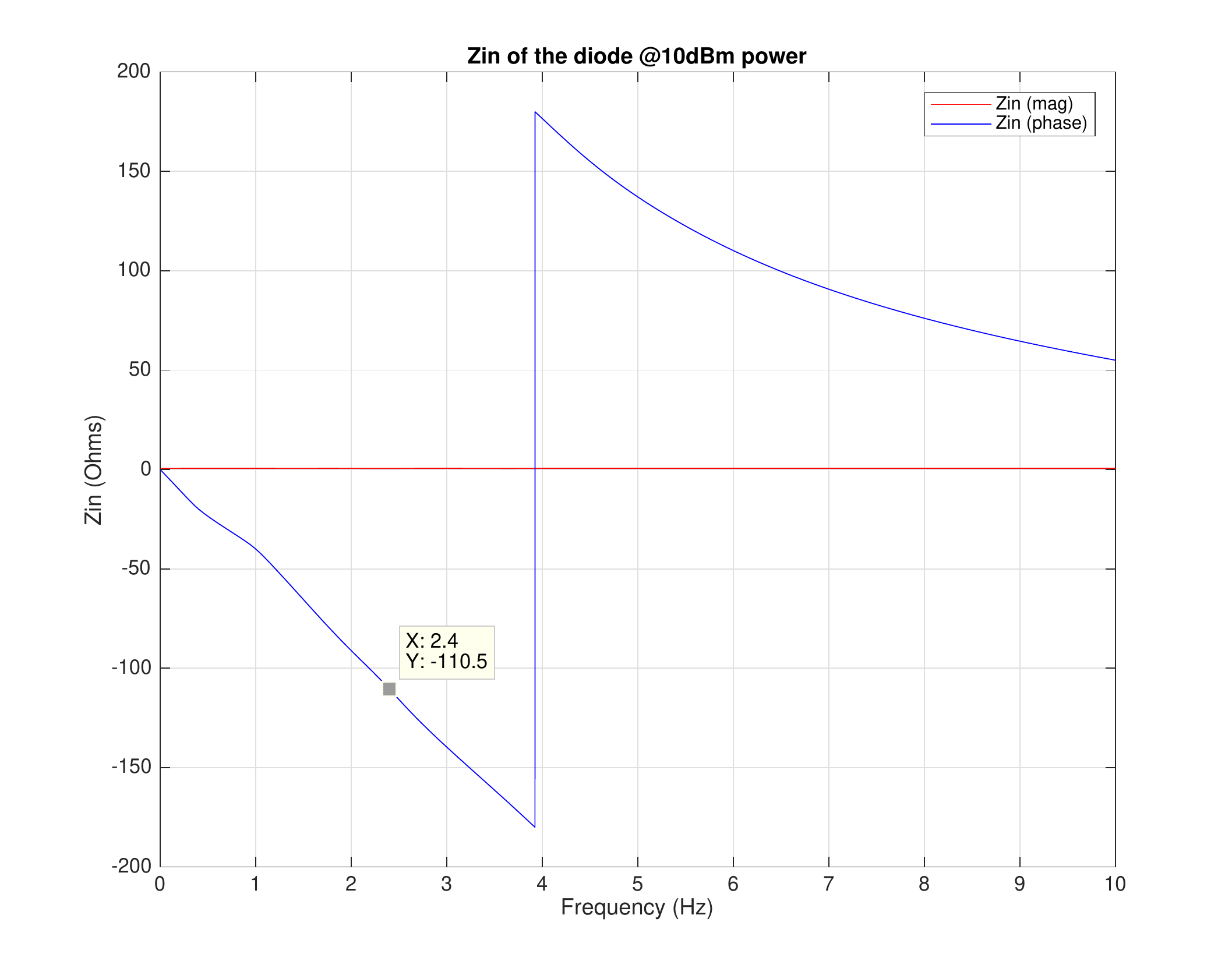}
\caption{$Z_{in}$ values of the diode in ADS}
\end{figure}\\

In the same way we compute the $Z_{in}$ of the diode in reverse bias case, since in our model the diode are placed in opposite directions. The $Z_{in}$ for opposite direction of the diode is 0.621-j*0.354. This $Z_{in}$ is the normalized one with $Z_{0}$ of 50 Ohms. \\

\subsection{Matched network for diode}
From the previous steps, we got the load impedance ($Z_{in}$) which gives the matched circuit need to design for rat-race coupler to the diode. Now we need to design the matched network for the source $Z_{0}$ and and load-impedance $Z_{in}$. The ADS software with smith chart utility has the the flexibility to design the matched network. \\

In this step, we designed load-impedance matched network for diode in forward bias. The load-impedance is 19.3-j*31.5 ohms. The following figure shows the load-matched network.
\begin{figure}[h!]
\centering
\includegraphics[width=8cm,height=3cm]{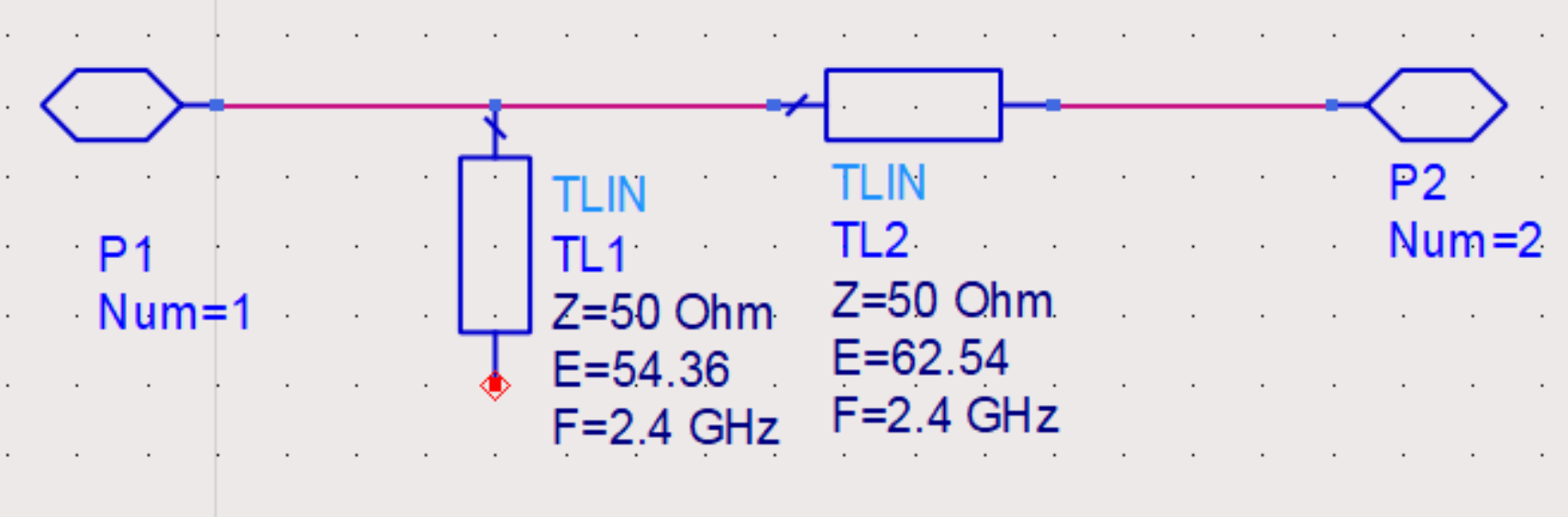}
\caption{Load-matching network from smith chart tool}
\end{figure} \\
Using the Smith Chart Utility, we got the schematics with equivalent electrical lengths. By using the Line-cal tool in ADS, we got the equivalent physical length of the transmission lines. We used open stubs which is easy to design, since these don't need to connect to via.
\begin{figure}[h!]
\centering
\includegraphics[width=8cm,height=3cm]{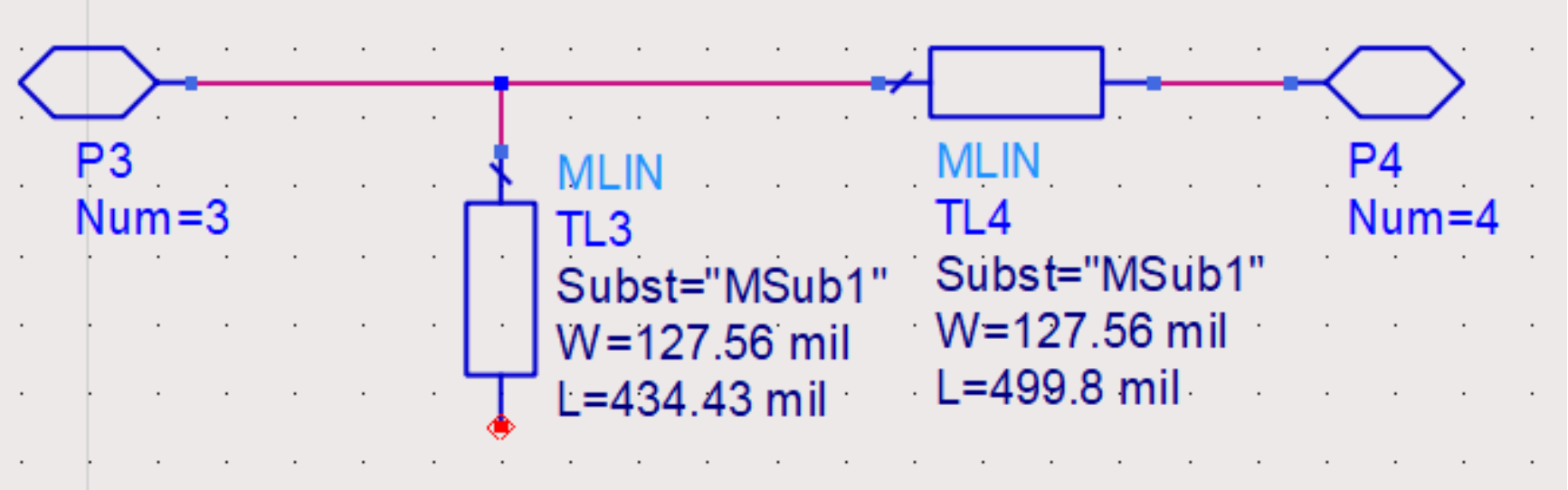}
\caption{Load-matching network with stub physical length}
\end{figure} \\
The Line-Cal tool gives us the length and width of the transmission lines for the specified impedance and Transmission lines length. In the same way, we computed the electrical and physical lengths of the diode matched network in reverse bias. The following figure shows the matching network for diode in opposite direction.
\begin{figure}[h!]
\centering
\includegraphics[width=8cm, height=3cm]{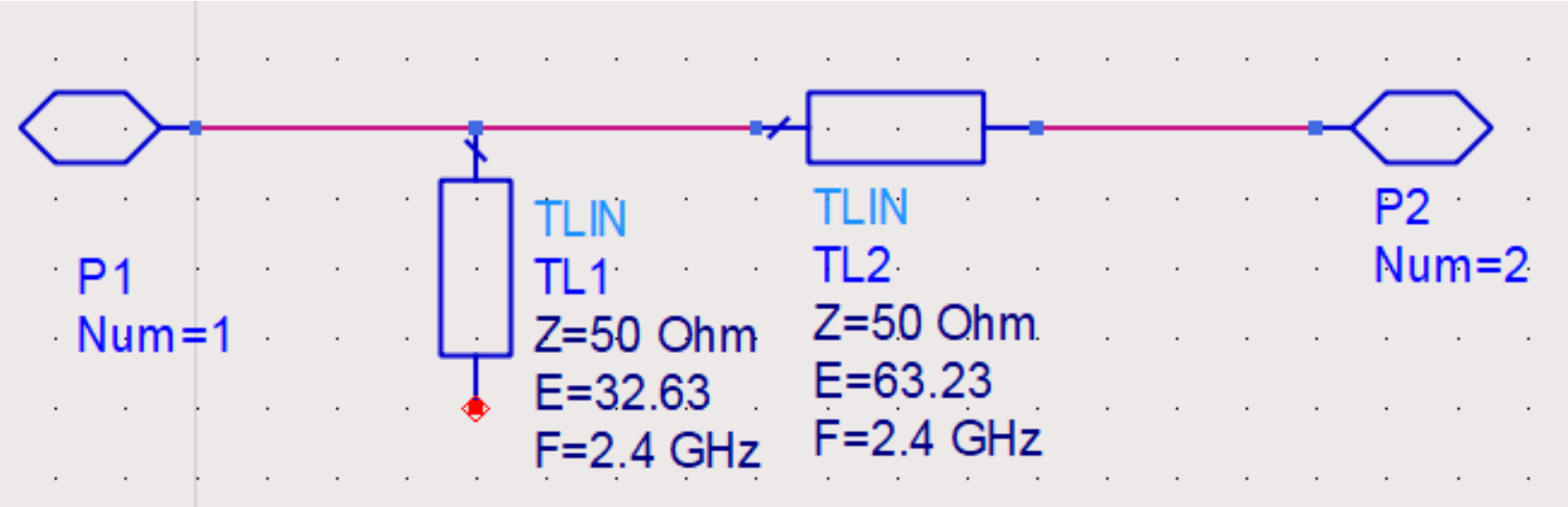}
\caption{Load-matching network for reverse direction of diode}
\end{figure}\\
The following figure shows the equivalent physical lengths of the matched network.  
\begin{figure}[h!]
\centering
\includegraphics[width=8cm,height=3cm]{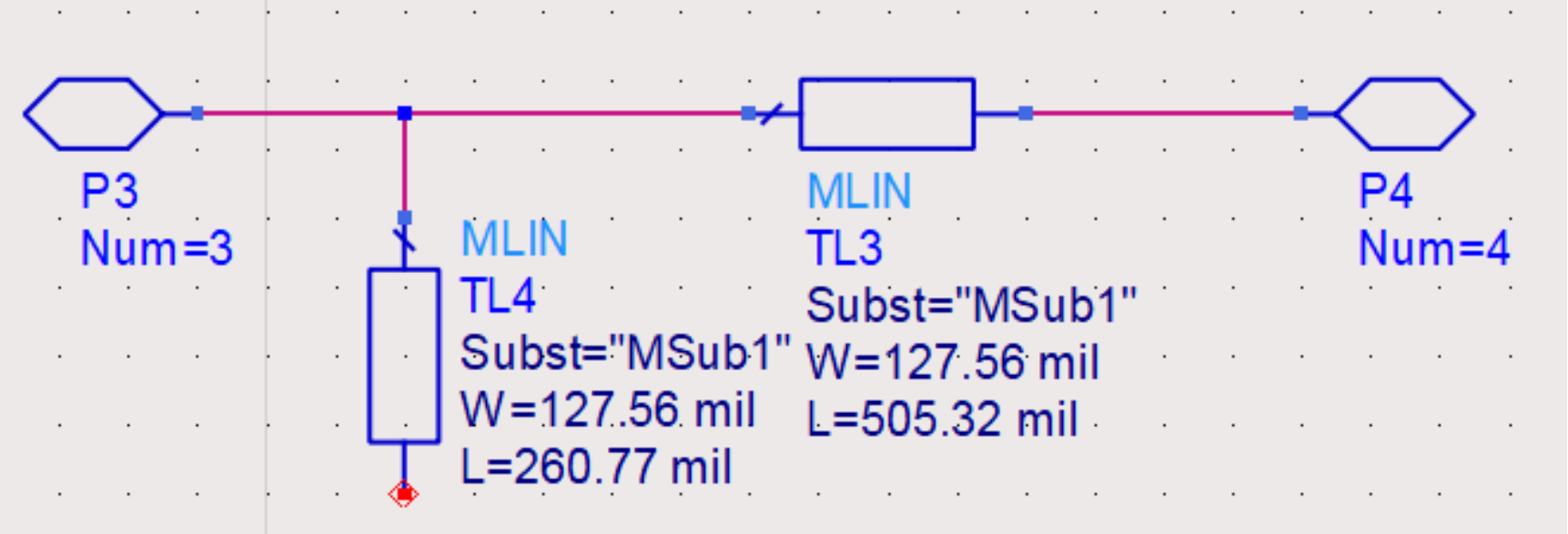}
\caption{Load-matching network with physical length for reverse direction of diode}
\end{figure}\\
Till this section, we finished the diode model and matched network of the diode. \\

\subsection{Rat-race Coupler}
The important component in the mixer design is the hybrid of 90 deg or 180 deg. Here we are using 180 deg hybrid which is the rat-race coupler. The rat-race coupler is having length of 1.5*$\lambda$ of length. Out of which 3 ports are separated by $\lambda/4$ and other one is separated by three times $\lambda/4$. The following figure shows the general rat-race coupler design.\\
\begin{figure}[h!]
\centering
\includegraphics[width=4cm,height=4cm]{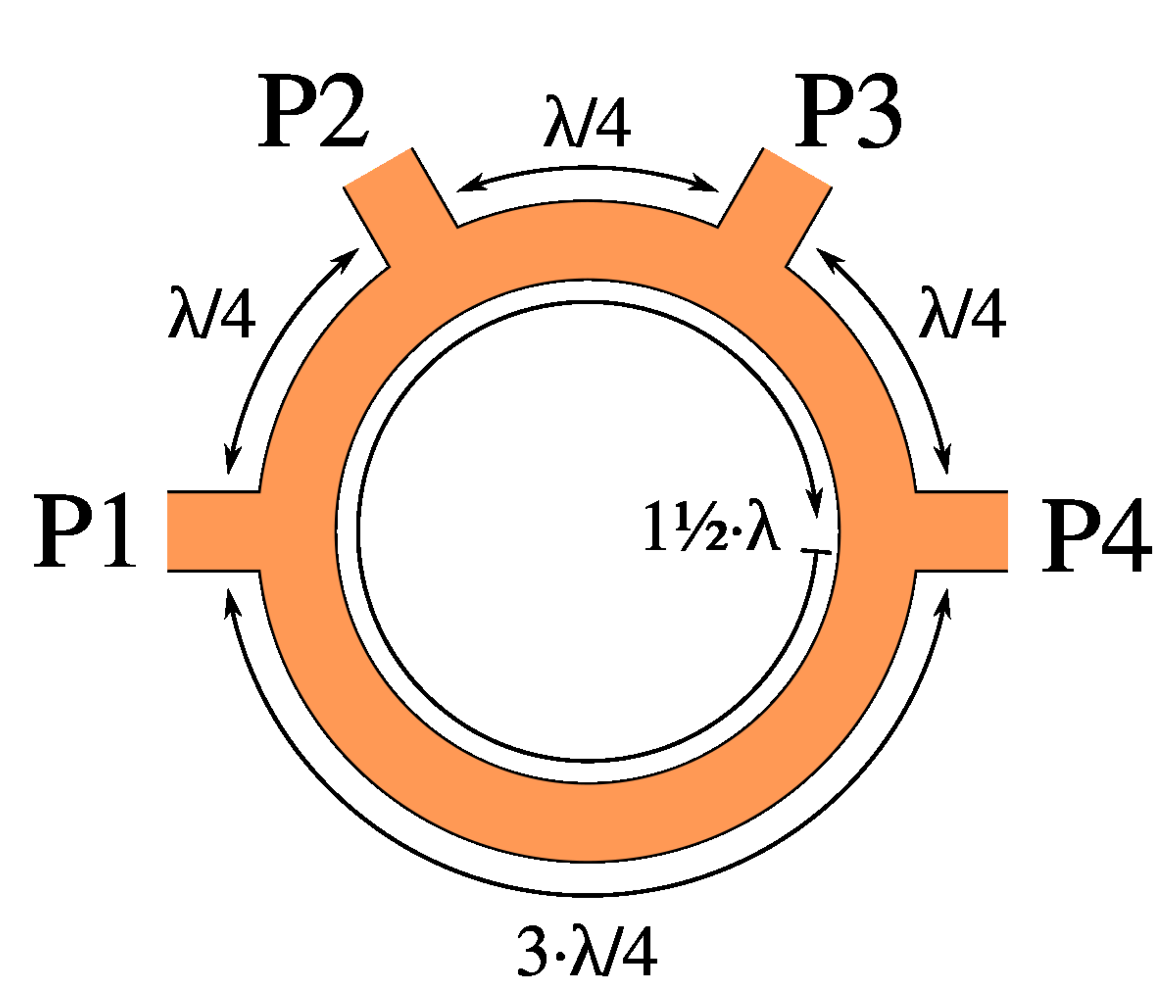}
\caption{Rat-race coupler}
\end{figure}
From the figure, the port P1 is the RF input, port P2 is the sum port, port P3 is the LO input and port P4 is the IF output port. The S-parameter matrix of the rat-race coupler is given below.\\
\begin{figure}[h!]
\centering
\includegraphics[width=2cm,height=2cm]{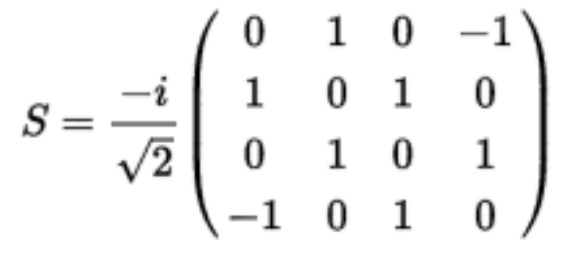}
\caption{S-matrix of the rat-race coupler}
\end{figure}\\
In this section, we designed the same rat-race coupler using ADS software.
Design steps:
Instead of $\lambda/4$ lines, we used the curves of the angel of $60^\circ$. Out of the 2-designs provided by the professor we are using design 1 to make the rat-race coupler.
The following figure shows the design 1 of the rat-race coupler.  
\begin{figure}[h!]
\centering
\includegraphics[width=7cm,height=5cm]{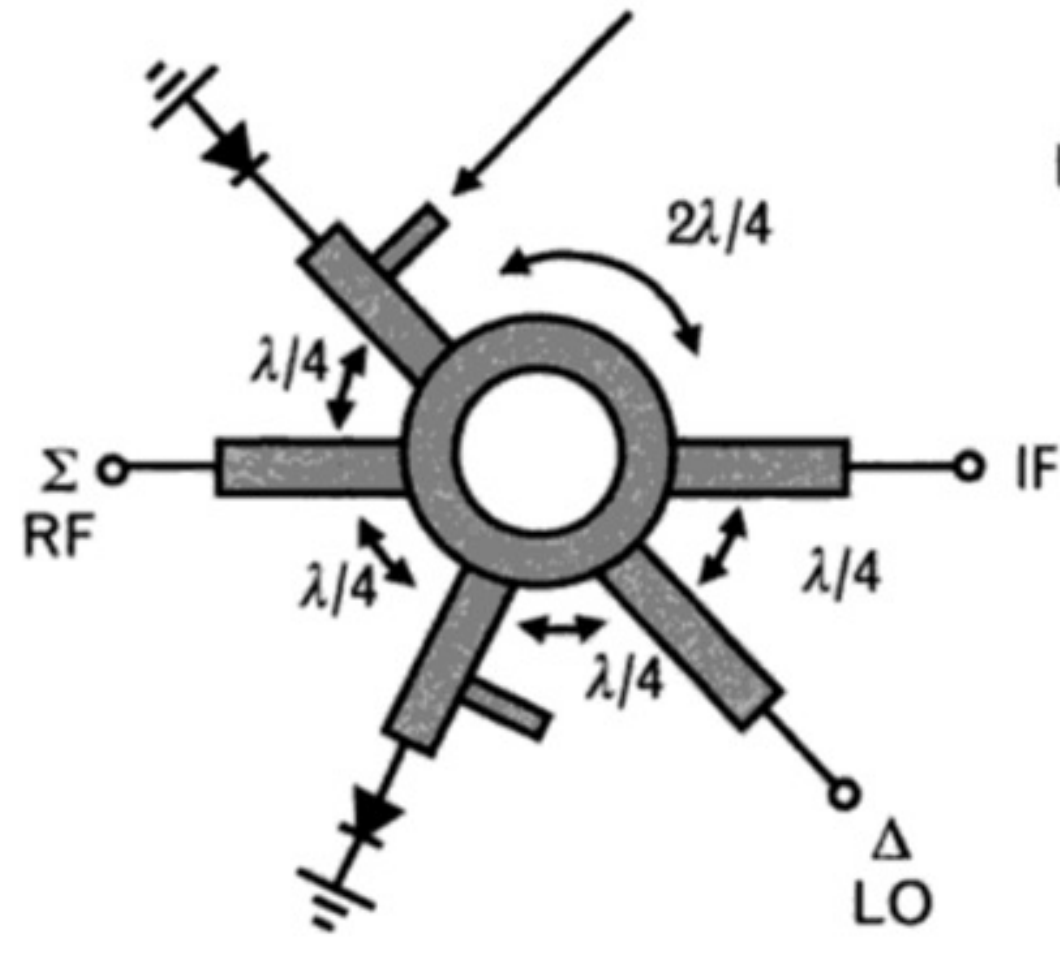}
\caption{chosen Rat-race design 1}
\end{figure}\\

As mentioned earlier, we are using curve of $60^\circ$ with same electrical length instead of $\lambda/4$ lines to make the design looks like a circle. We added another $\lambda/4$ division so that the diode can be placed symmetric from $3*\lambda/4$ section. The "\textit{MCurves}" are connected using matched Tee sections. To make the diode and LO ports vertical , we used another $30^\circ$ curve before connecting the ports. 
\begin{figure}[h!]
\centering
\includegraphics[width=9cm,height=5cm]{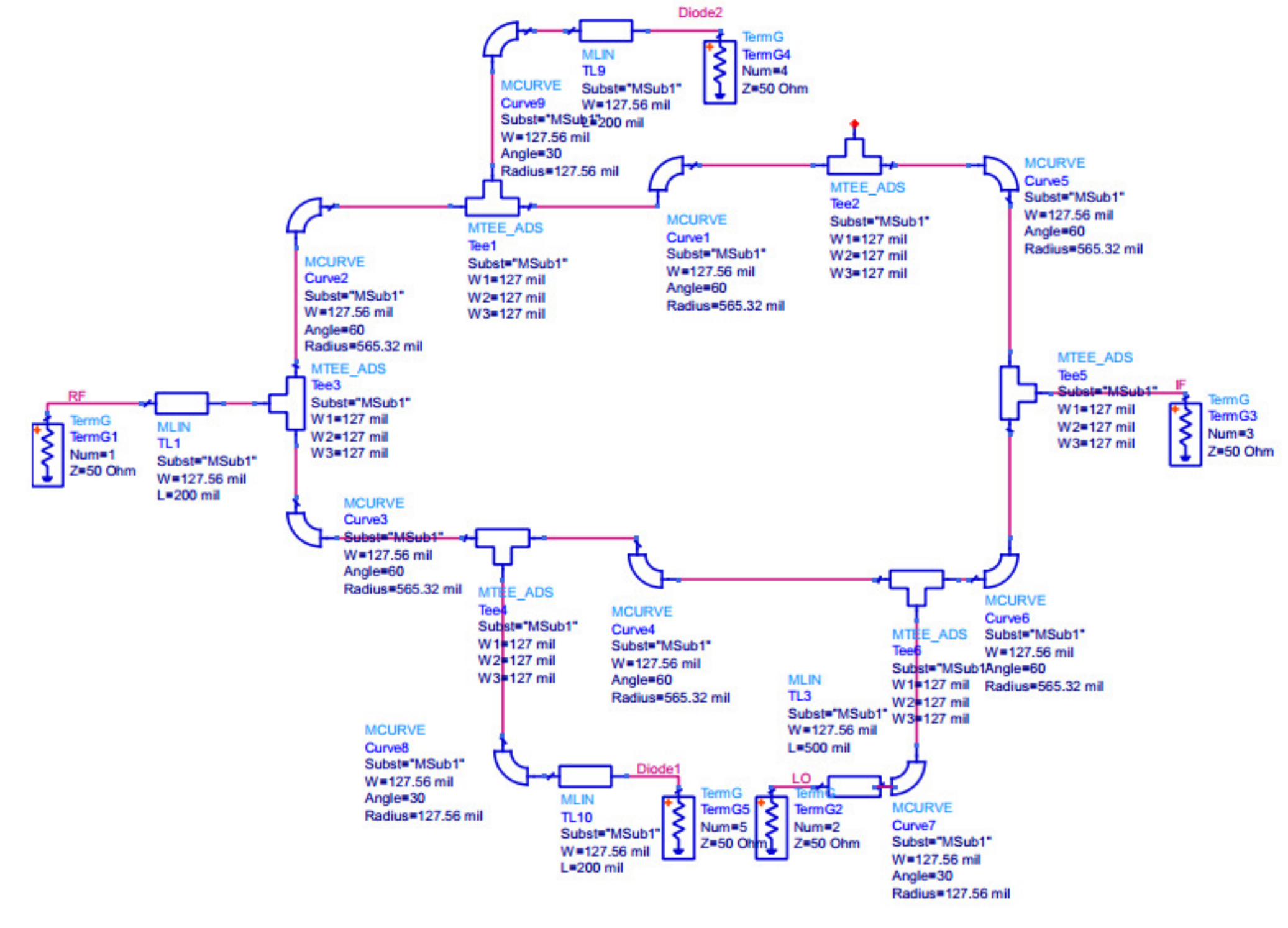}
\caption{Rat-race schematic}
\end{figure} \\
Now we connected the matched network and diode equivalent model which are obtained in the earlier steps. By doing this we can suppress the LO signal. The diodes one end is connected to the rat-race coupler where as other end is connected to the ground. The ground is modeled as "VIAGND" component in ADS. The following figure shows the rat-race hybrid with the diodes.
\begin{figure}[h!]
\centering
\includegraphics[width=9cm,height=7cm]{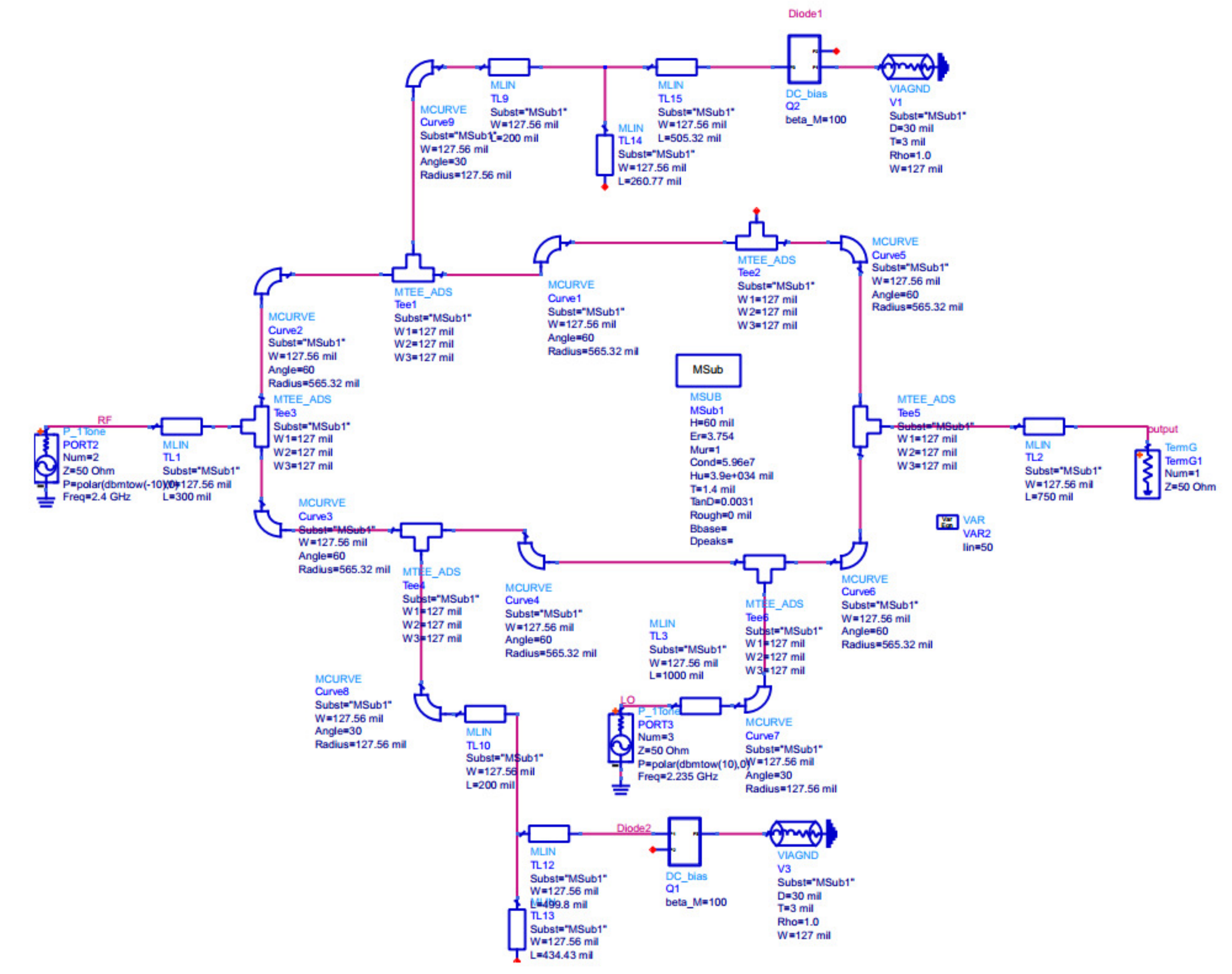}
\caption{Rat-race schematic with diodes}
\end{figure} \\

From the figure, the diode are placed in opposite directions. We added the matched network of the diodes before connecting them to the rat-race coupler. The following figure shows the S-parameters of the rat-race coupler. 
\begin{figure}[h!]
\centering
\includegraphics[width=9cm,height=5cm]{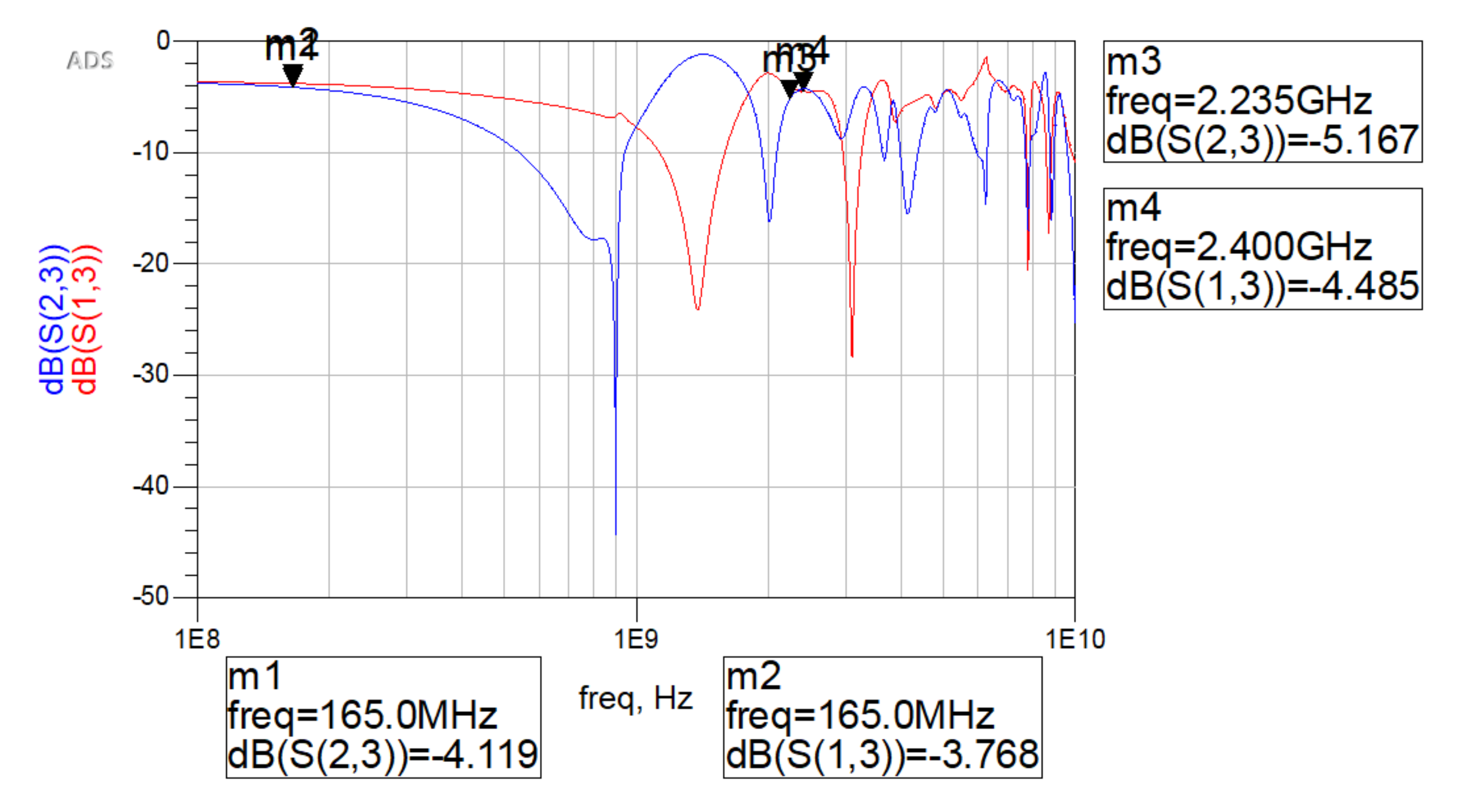}
\caption{S-parameters of the rat-race with diodes}
\end{figure} \\

Here the RF is at port1, LO is at port2 and IF at port3. clearly the LO is the high power signal than RF and results at the IF output. It is required to get only IF frequency at the IF. So we pass the IF signal with inter-mod products through the low-pass filter (IF filter) to suppress other frequency components. \\

\subsection{IF filter design}
From the previous section, it is clear that we have to suppress all other non-linear inter-mod products expect at IF frequency. So we defined the IF as below. \\

Definition: Passband ripple is less than 0.1dB, passband frequency 300 MHz, stop-band attenuation is less than 80dB, stop-band frequency will contain IM products of all orders (here we took at 1.2GHz to keep the filter order less and wide transition band). \\ 

In this section, we used the butterworth filter which is maximally flat in the pass band with order of 6 using \textit{filter design smart component}. We tuned the lumped components little bit so that it will be exact match with real component values using \textit{Tuning} option in ADS. The following circuit shows the IF filter used in the design. We used a coupling capacitor for blocking dc component from IF.
\begin{figure}[h!]
\centering
\includegraphics[width=7cm,height=4cm]{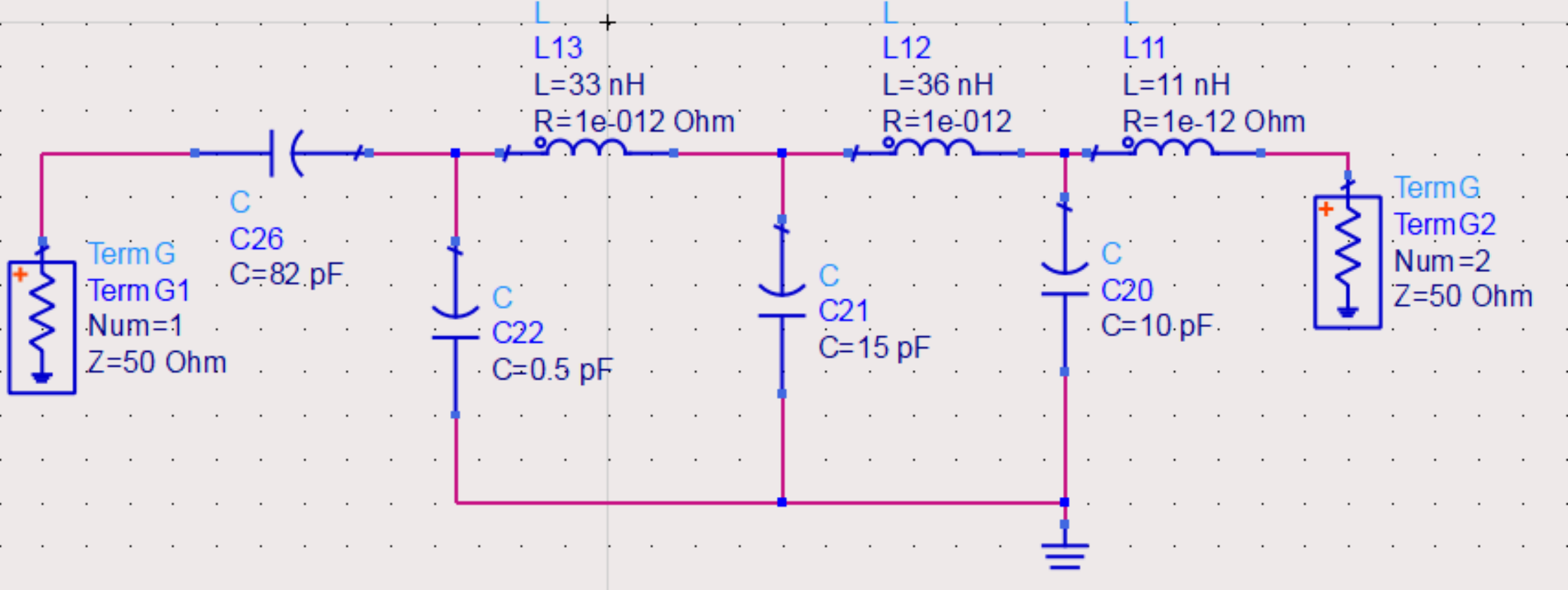}
\caption{IF filter schematic}
\end{figure} \\

We made the physical realization of the each component using their S-parameters in ADS. For example, the following figure shows the physical realization of 11nH inductor.
\begin{figure}[h!]
\centering
\includegraphics[width=7cm,height=4cm]{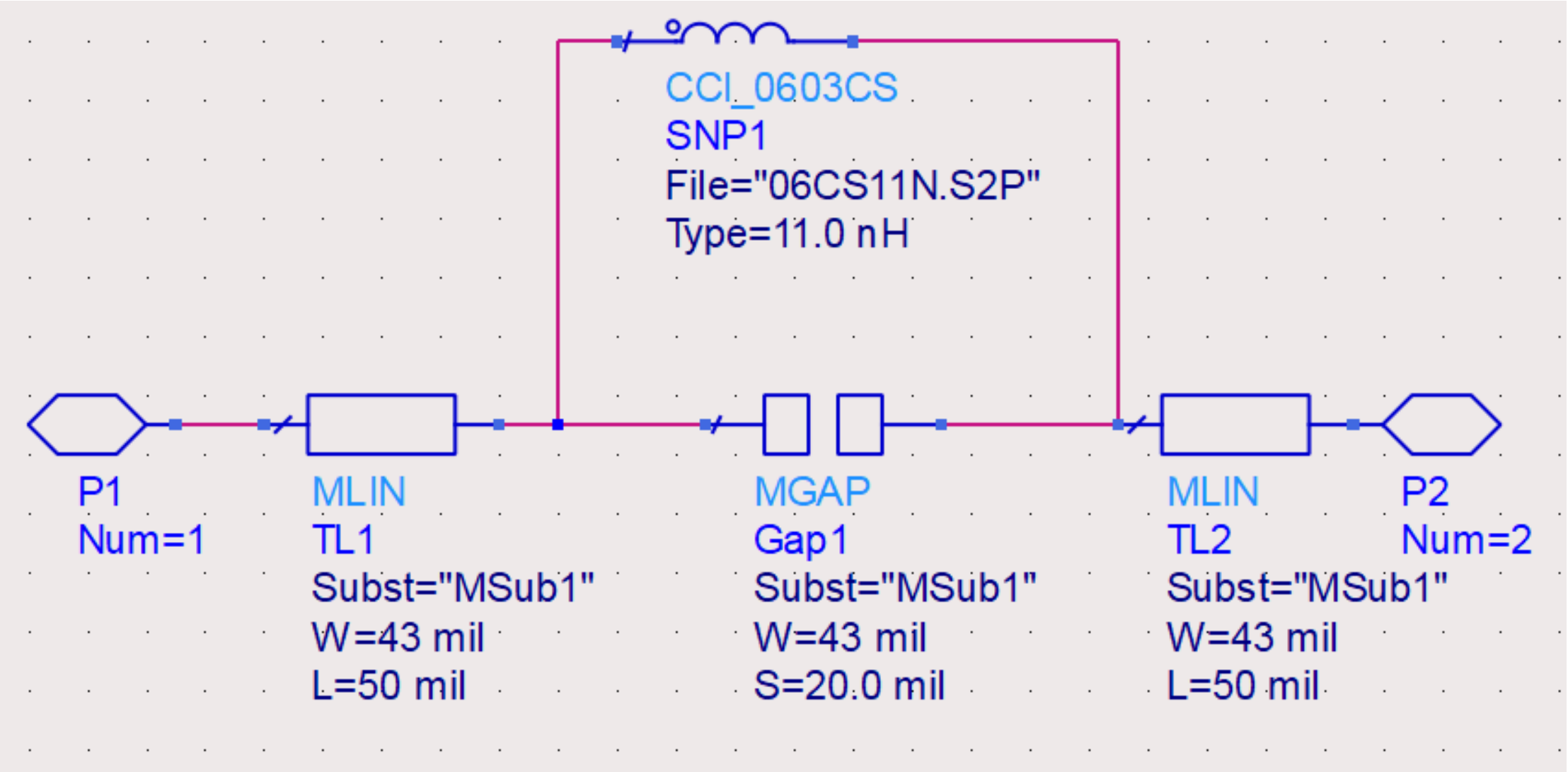}
\caption{IF filter schematic}
\end{figure} \\

The symbol is created for this schematic, so that we can use later. We did the same for all lumped elements. We used the matched "Tee" sections to connect these lumped elements. We used the tee section each of width is 43 mils. For inductors we are using 0603CS package.
\begin{figure}[h!]
\centering
\includegraphics[width=8cm,height=6cm]{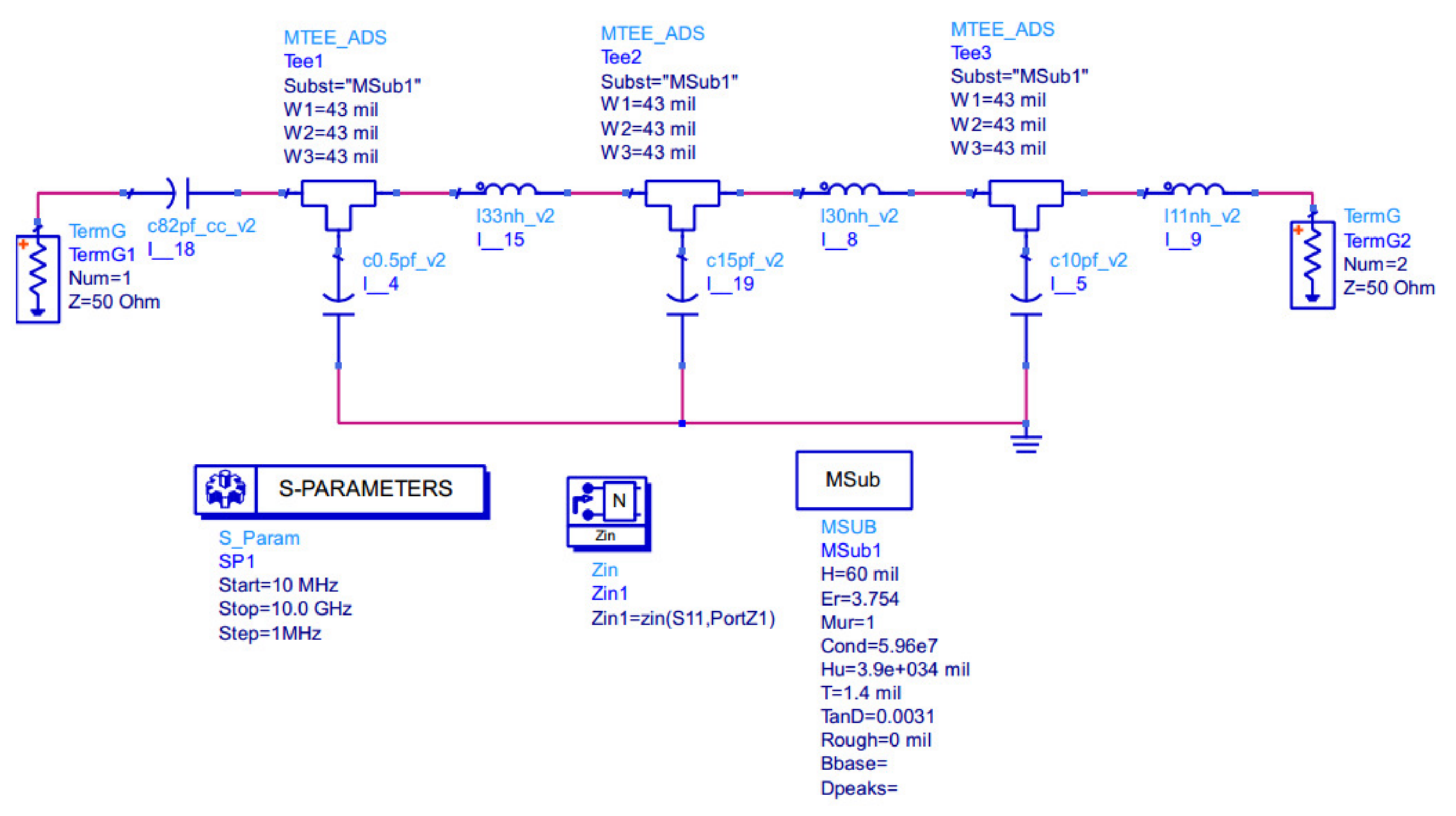}
\caption{IF filter with magic tees}
\end{figure}\\

The following figure shows the S-parameters of the IF filter. Clearly from this figure, the LO and RF signal are attenuated by more than 80dB.
\begin{figure}[h!]
\centering
\includegraphics[width=8cm,height=7cm]{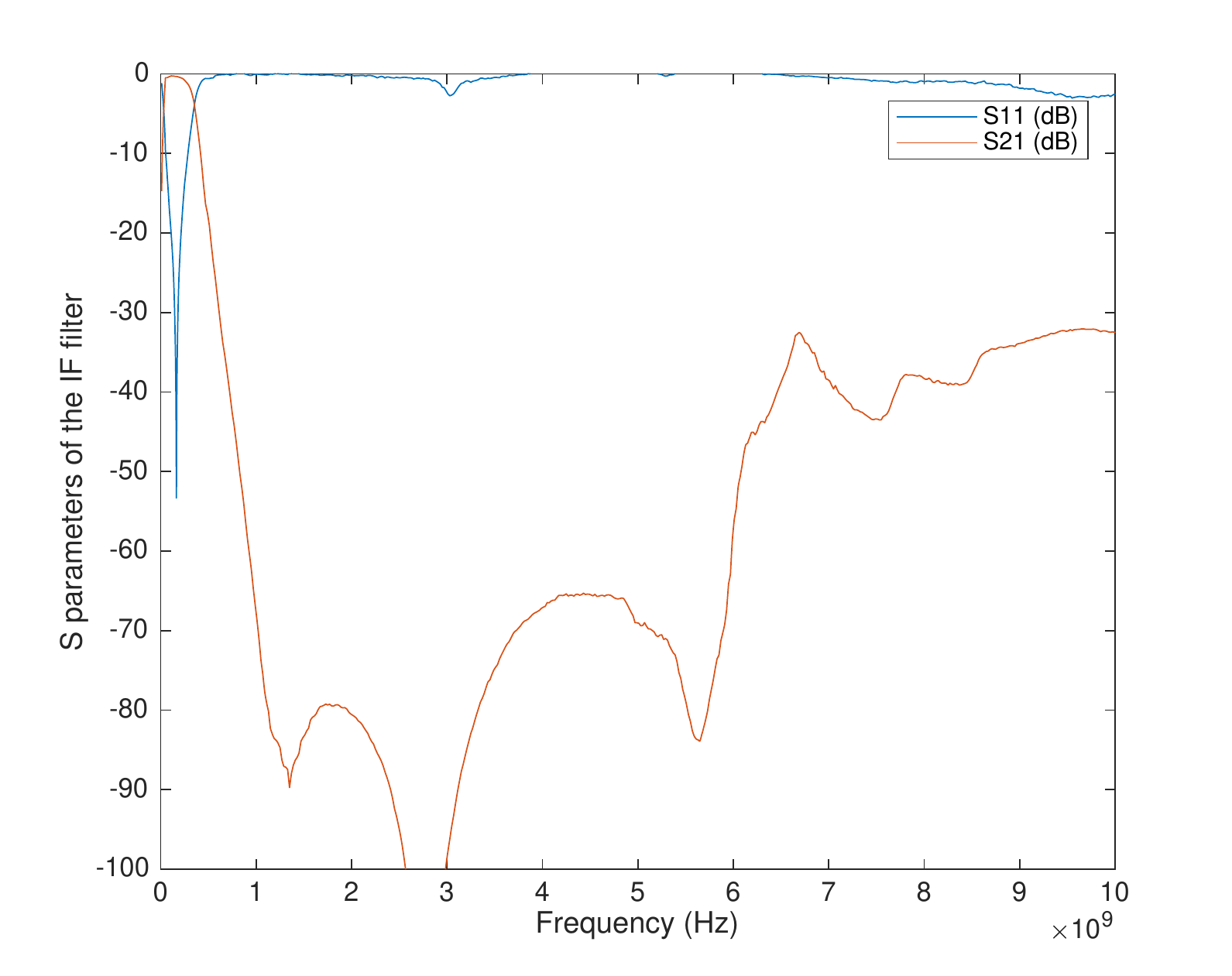}
\caption{S-parameters of the IF filter}
\end{figure} 
The zoomed version of IF filter is shown below. (till 2.5GHz)
\begin{figure}[h!]
\centering
\includegraphics[width=8cm,height=7cm]{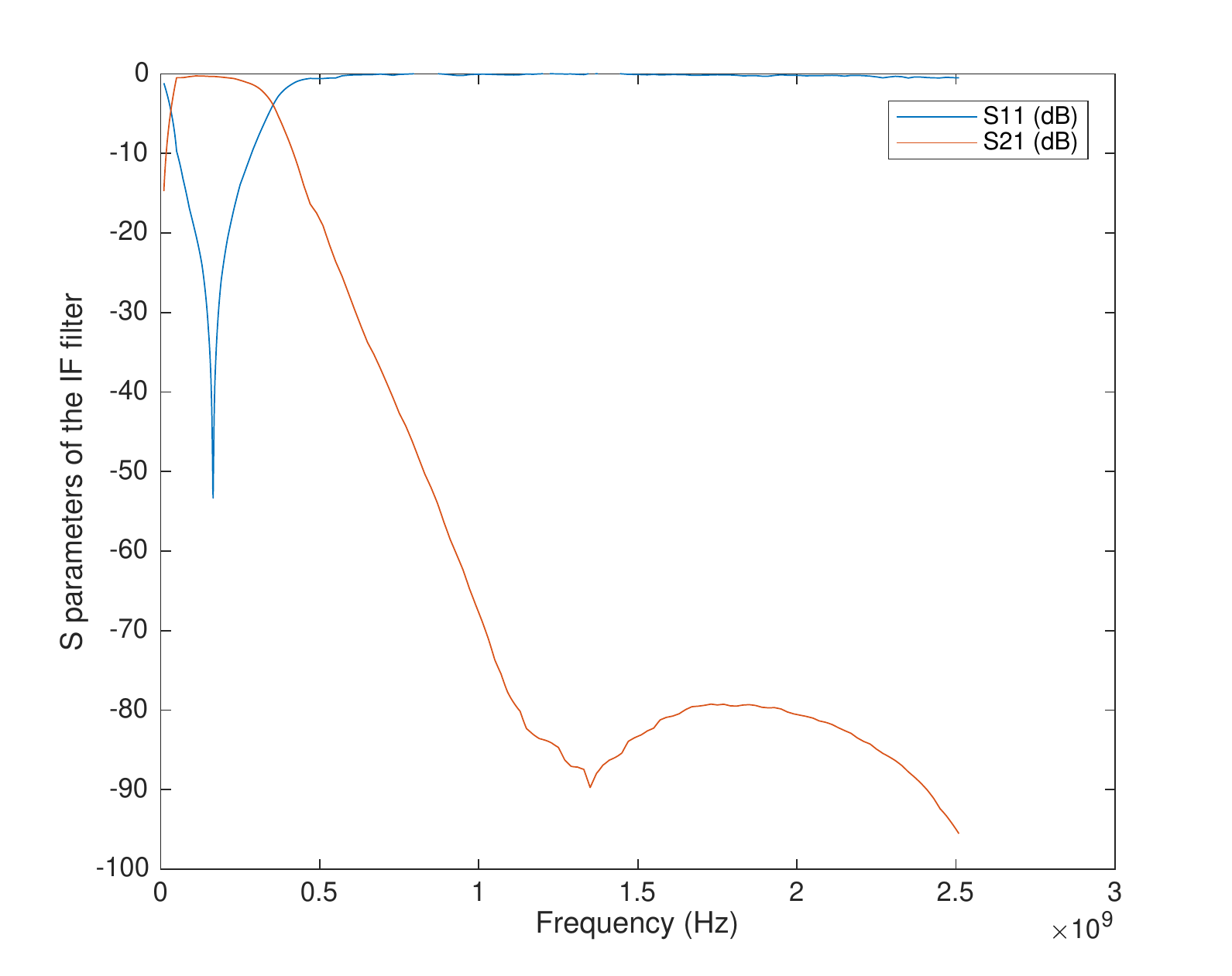}
\caption{S-parameters of the IF filter}
\end{figure} 

\subsection{Complete Mixer}
In this section, we added the rat-race coupler with the IF filter. The following figure shows the ADS schematic of the rat-race coupler with IF filter.
\begin{figure}[h!]
\centering
\includegraphics[width=8cm,height=5cm]{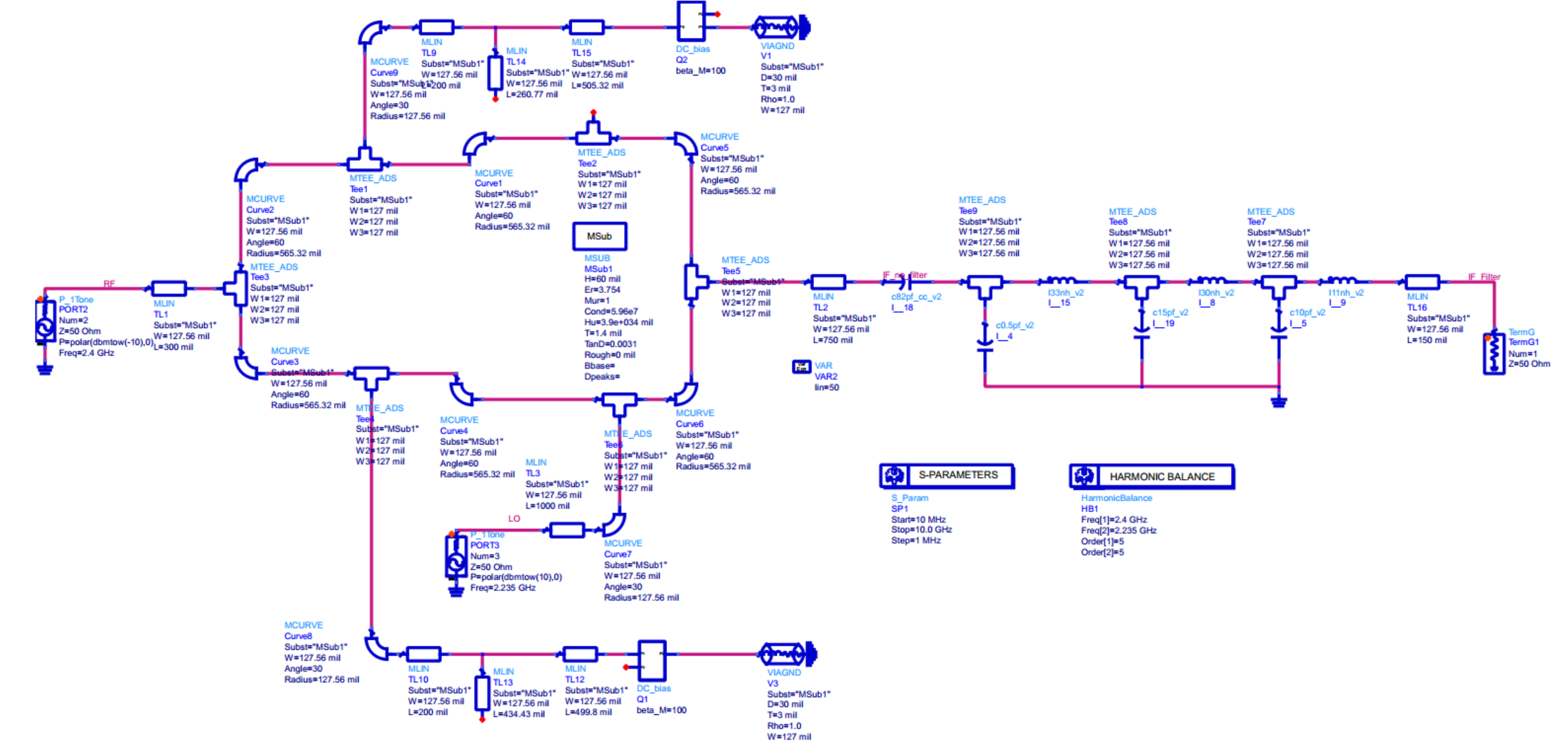}
\caption{Mixer schematic}
\end{figure} \\
The following figure shows the IF spectrum of the mixer circuit. 
\begin{figure}[h!]
\centering
\includegraphics[width=9cm,height=6cm]{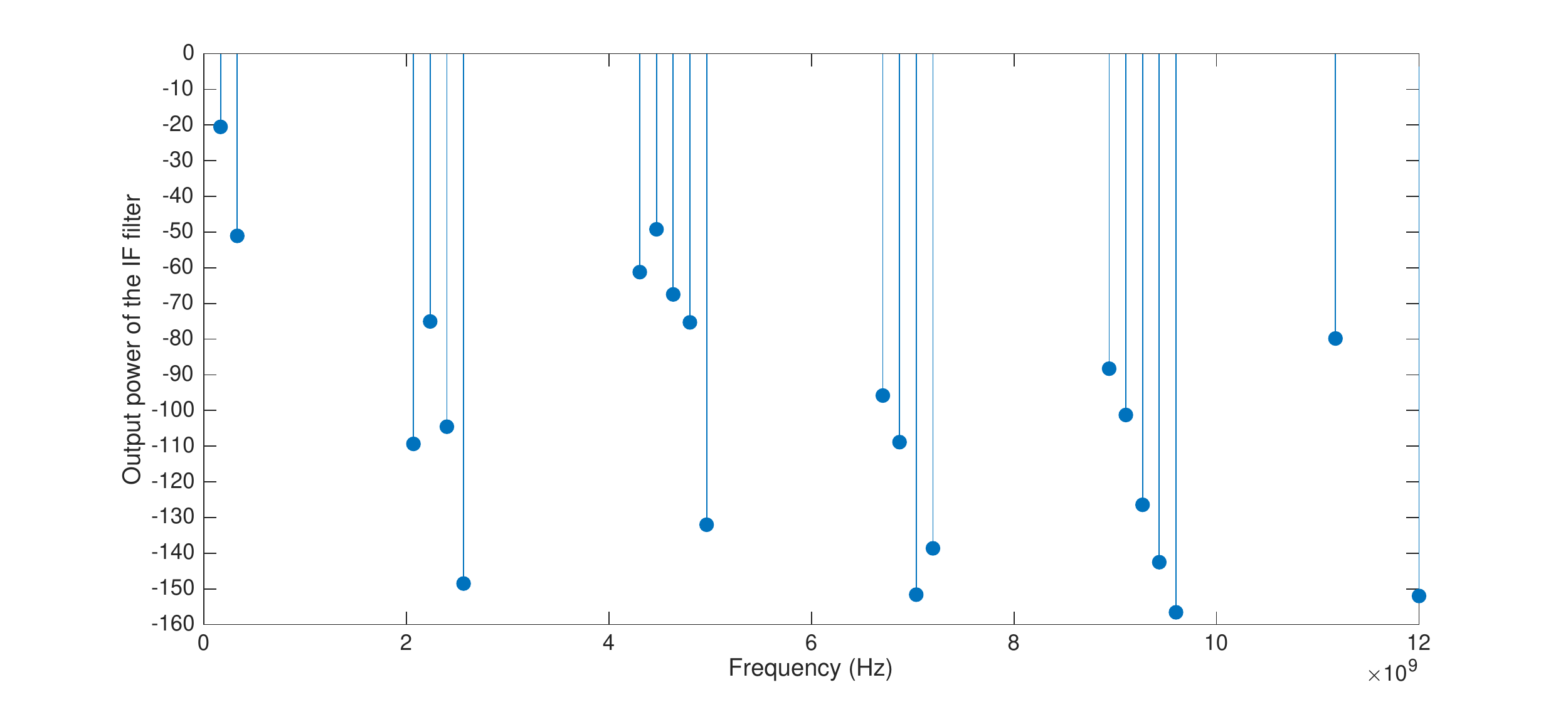}
\caption{Output products of the mixer circuit from Matlab}
\end{figure} 
From the above figure, it is clear that RF to OF loss is less than 8dB, LO signal power is given by -75 dBm and RF signal at IF port is at -104 dBm. 
\begin{figure}[h!]
\centering
\includegraphics[width=8cm,height=6cm]{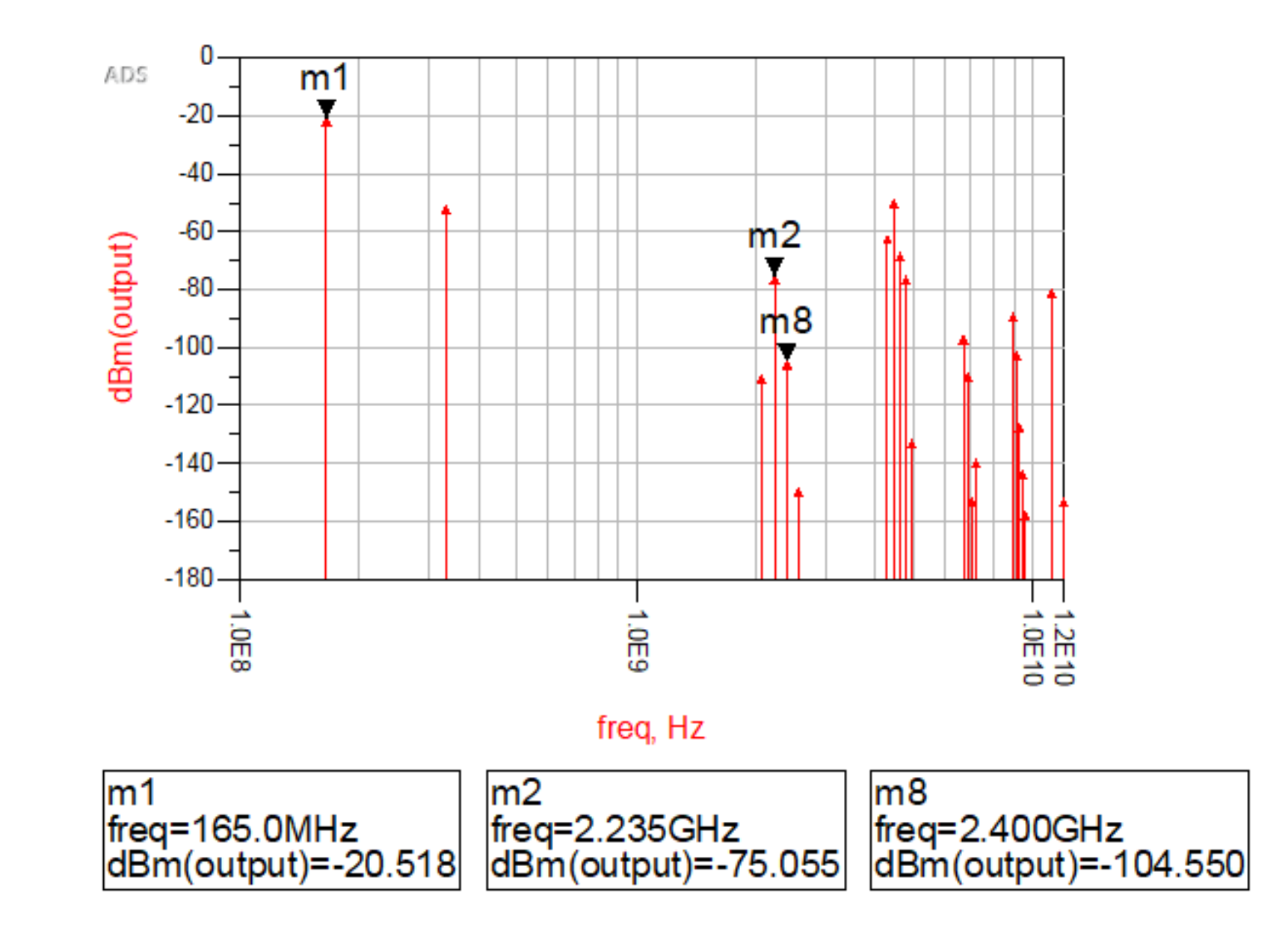}
\caption{Output products of the mixer circuit from ADS}
\end{figure}\\
The S-parameters of the mixer circuit is given in the following figures.
\begin{figure}[h!]
\centering
\includegraphics[width=8cm,height=6cm]{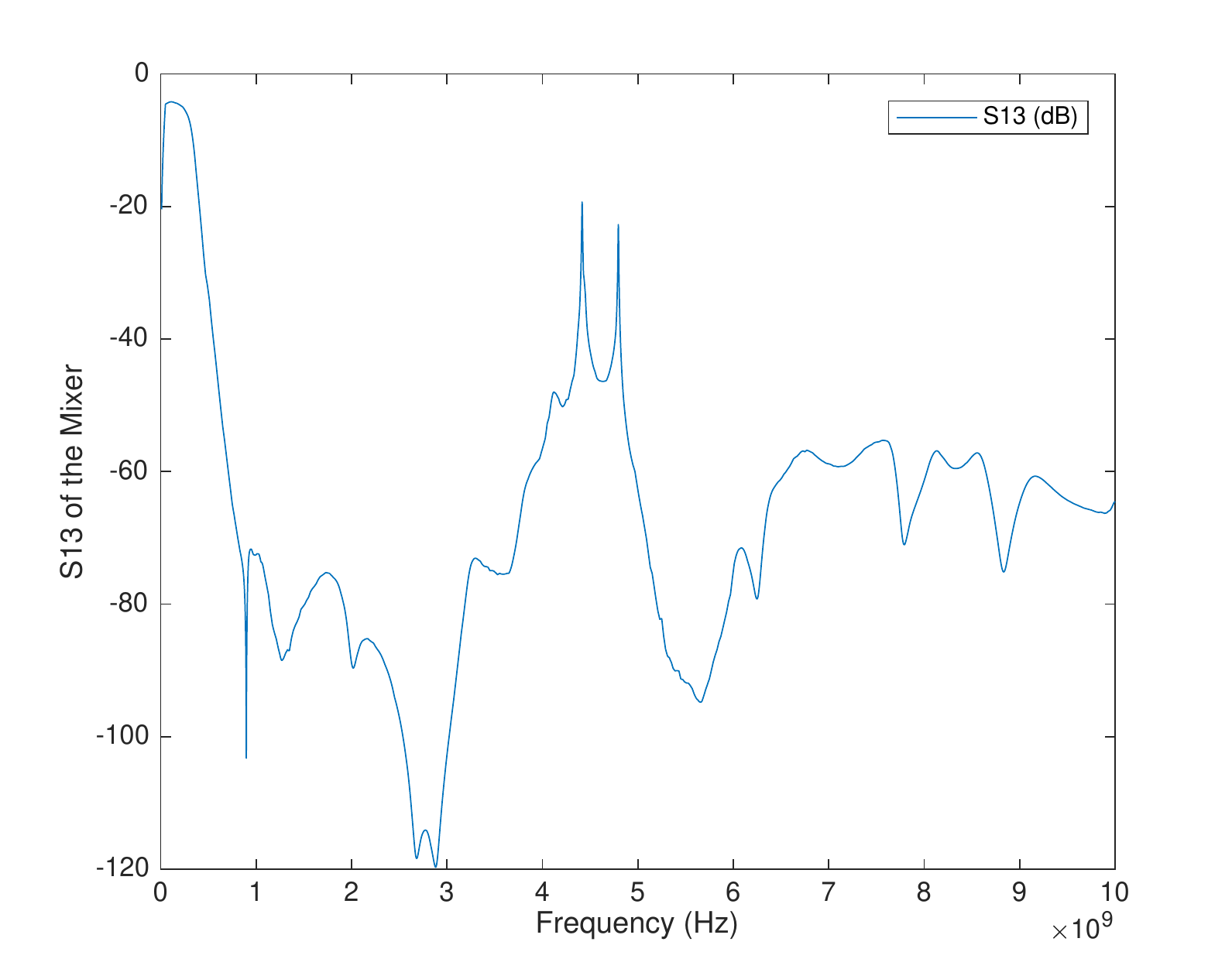}
\caption{RF to IF S-parameters}
\end{figure}
\begin{figure}[h!]
\centering
\includegraphics[width=8cm,height=6cm]{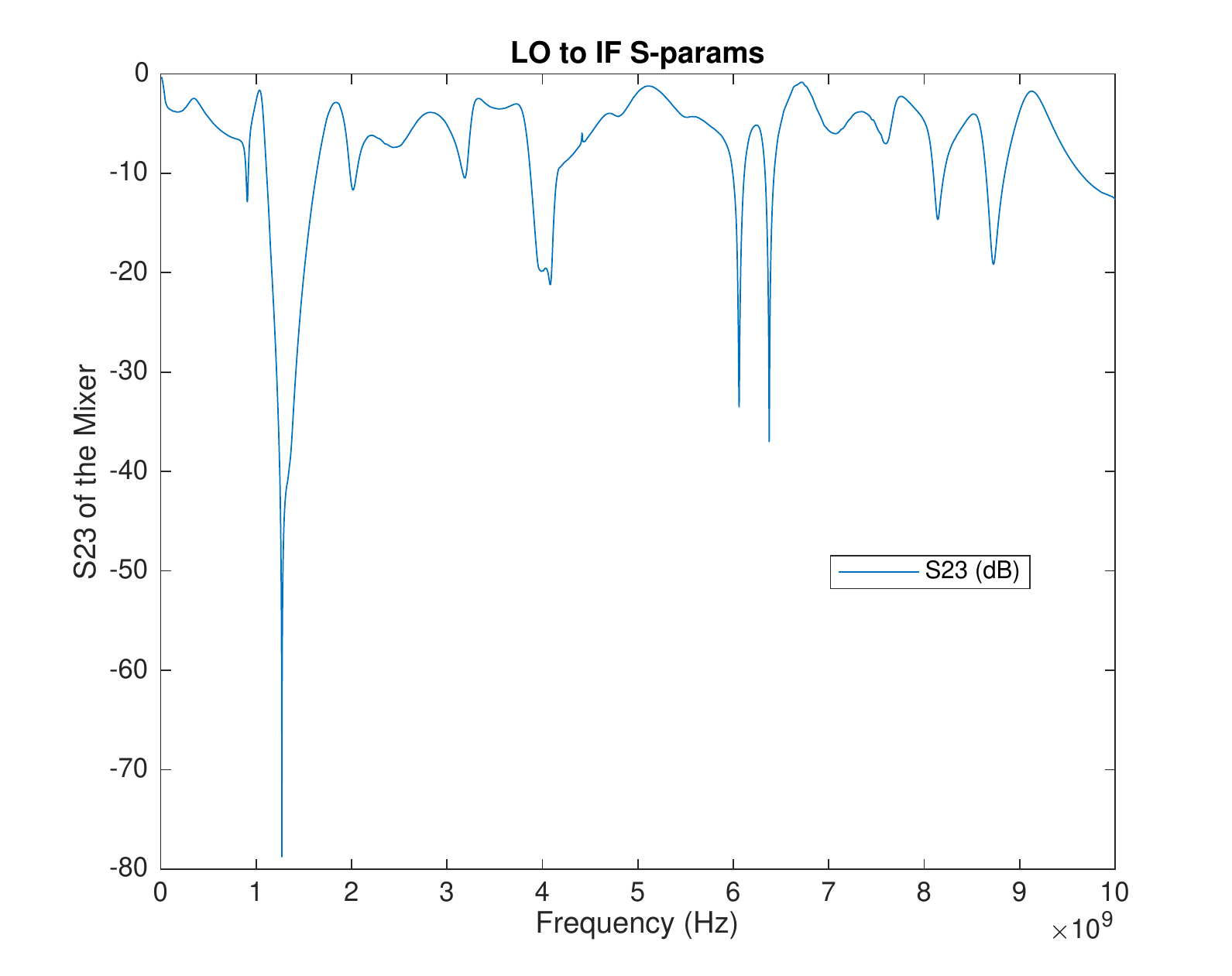}
\caption{LO to IF S-parameters}
\end{figure}
\begin{figure}[h!]
\centering
\includegraphics[width=8cm,height=6cm]{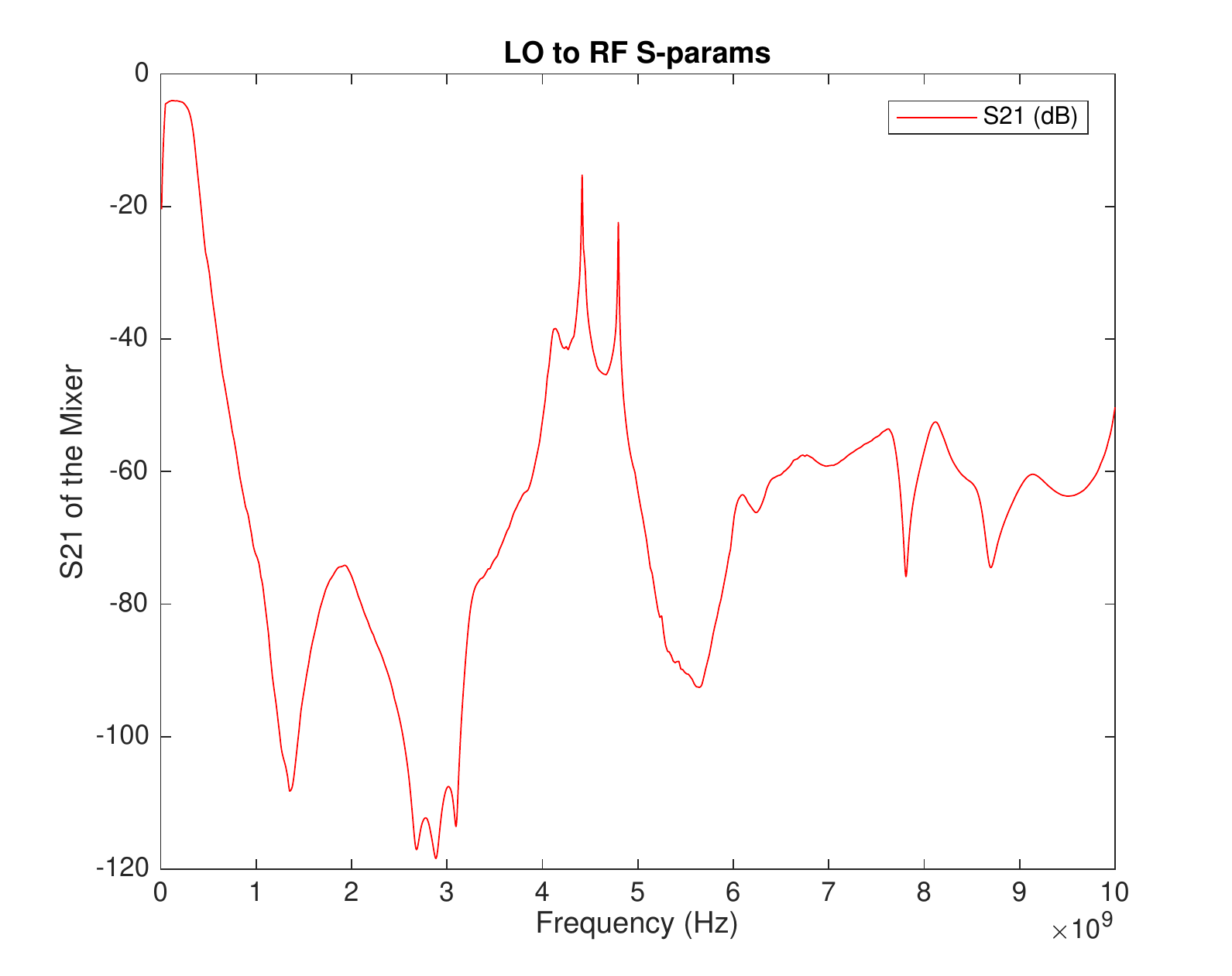}
\caption{LO to RF S-parameters}
\end{figure}

Clearly, from the S-parameters, the conversion loss is given by -4.43 dB which is RF to LO, RF to IF @RF frequency is given by -91.2 dB, the LO to IF isolation is given by -6.241 dB, and the LO to RF isolation is given by -84.92 dB. From this it is clear that the LO matching is not proper at the IF output. The matched network of the all ports are explained in the later sections. \\

\subsection{Layout design}
In this section, we replace all the schematics with appropriate layout circuits. The complete layout of the schematic is shown below.  
\begin{figure}[h!]
\centering
\includegraphics[width=8cm,height=8cm]{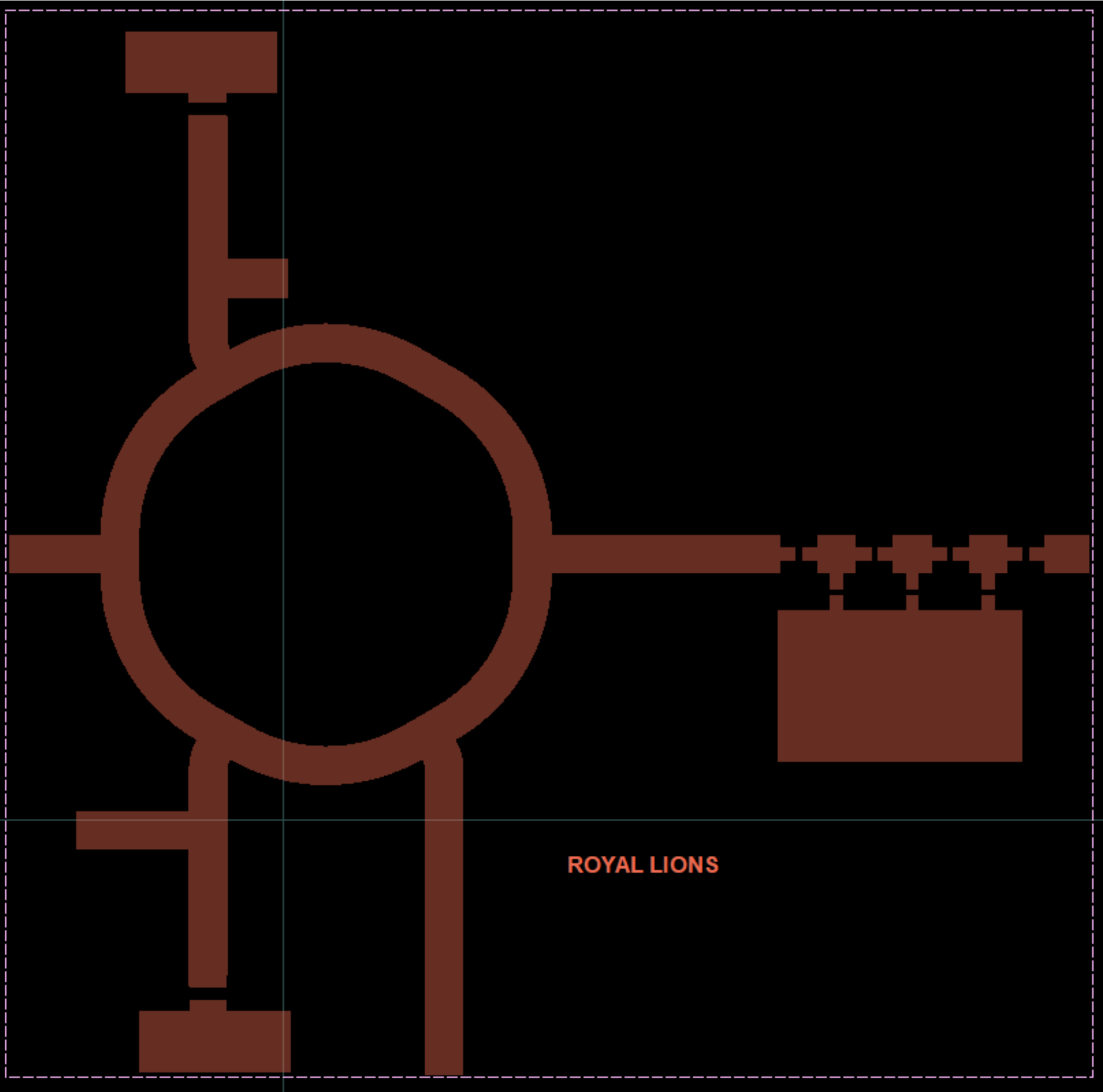}
\caption{Layout of the mixer circuit}
\end{figure}
The capacitors used in the design are 0.5 pF, 10 pF, 15 pF and 82 pF (coupling capacitor $C_c$). The inductors used in the design are 11 nH, 30 nH and 33 nH. In the schematic, additional transmission line of 200 mils is added to make sure we have enough transmission line to place the ports of the transistors at all the ports. We added \textit{viaholes} for short circuit. We dig the holes for short-circuit in the design.\\

\section{Fabrication}
The layout using Gerber file is fabricated using the LPKF machine. We used the Rogers 4350B substrate with effective resistance of 3.754, thickness is 60mil, 1oz copper (1.4 mil), loss tangent of 0.0031 and conductivity of 5.96e7. We selected the appropriate settings for the milling. The following figure shows after fabrication from LPKF machine. 
\begin{figure}[h]
\centering
\includegraphics[width=8cm,height=6cm]{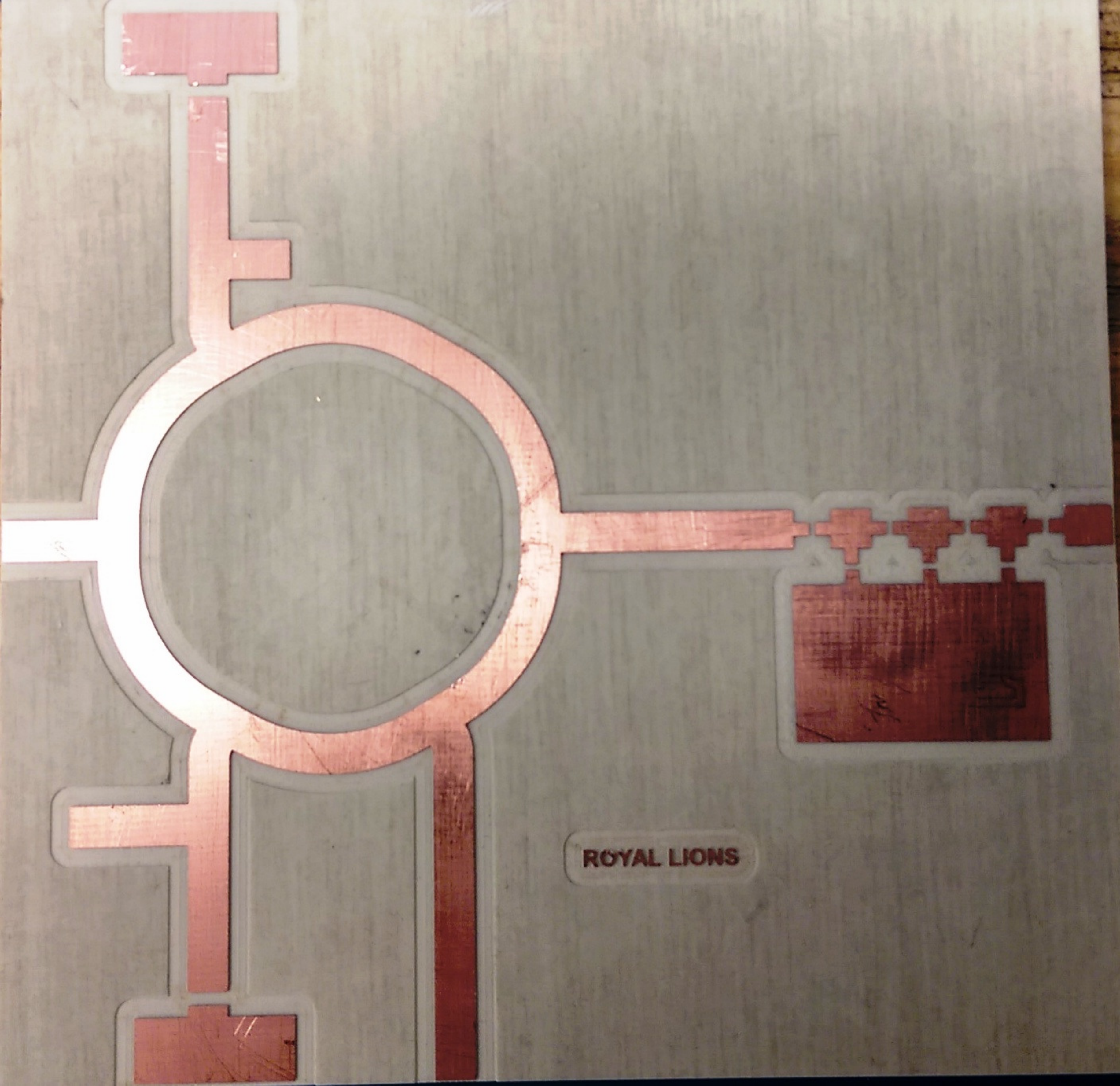}
\caption{Fabricated Mixer with LPKF machine}
\end{figure}\\
\subsection{Lumped components and diodes} 
We soldered the capacitors, inductors and diodes on the fabricated design. We added the 3-ports for RF input, LO input and IF output with 3.5mm SMA female connectors. 

The following figure shows after adding lumped elements and diodes.  
\begin{figure}[h]
\centering
\includegraphics[width=8cm,height=6cm]{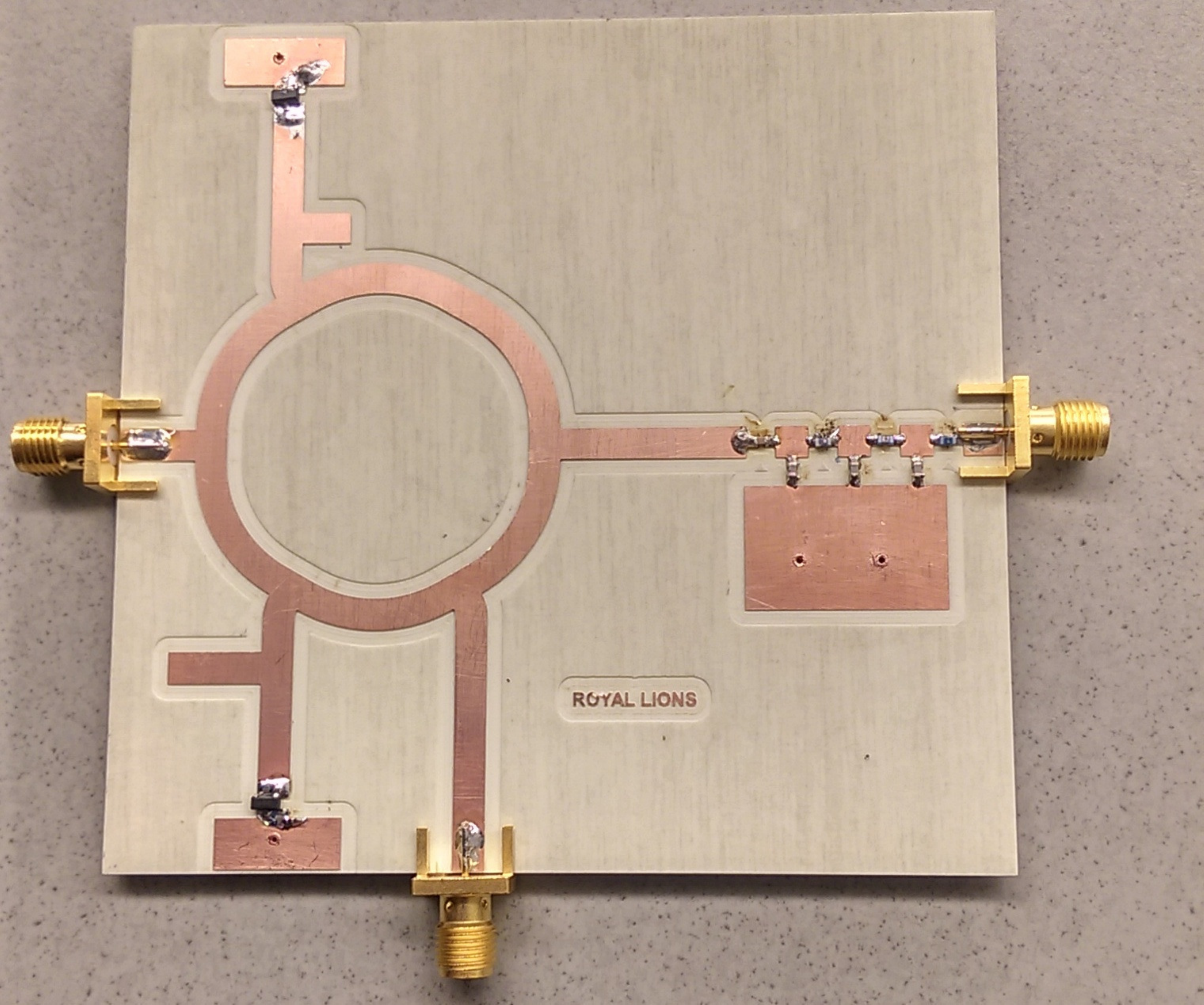}
\caption{Mixer after soldering the components}
\end{figure}\\

\section{Measurements}
The following are the performance metrics of the mixer.
\begin{enumerate}
\item Conversion loss: $C_L$= $P_{rf}$-$P_{if}$ which is opposite of the conversion gain ($C_G$)
\item Isolation of ports: isolation across all ports
\item 1-dB compression point: impact on switching of RF power
\item Noise figure: proportional to Conversion loss
\item single tone IM distortion: unwanted harmonics are called IMD
\item Multi-tone IM distortion: using 2-tone test,measure IIP3 value.
\end{enumerate} 
Out of which here we measured the conversion loss, isolation and IIP3 values of the designed mixer. The following figure shows the typical setup for mixer measurements.
\begin{figure}[h]
\centering
\includegraphics[width=8cm,height=3cm]{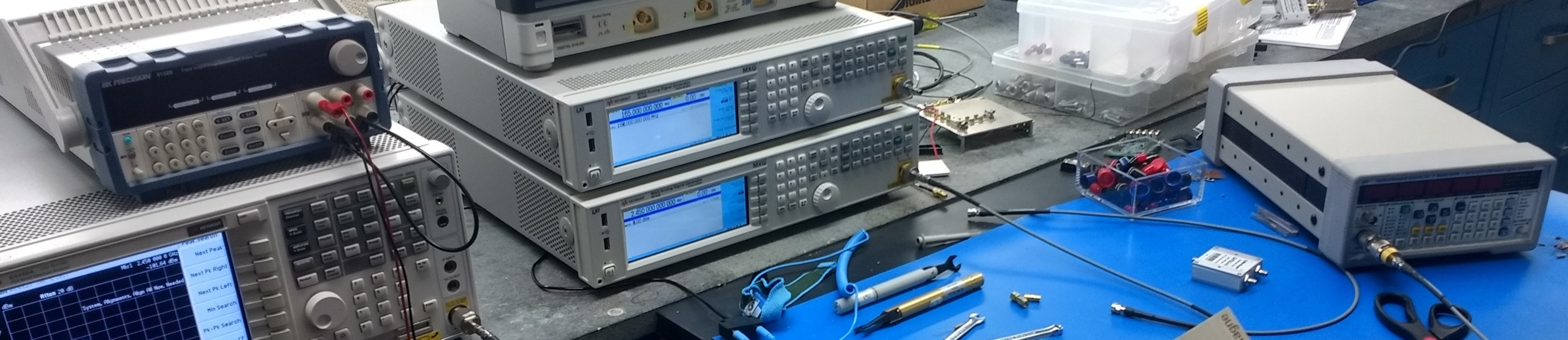}
\caption{Mixer measurement setup}
\end{figure}\\

\subsection{Calibration}
Before measuring the loss parameters, it is better to calibrate the cables and know the cables loss. Before inserting the Mixer in the measurement, perform the port1 and port2 loss which includes the splitter. We used the Keysight vector signal generator(s) and signal analyzers for the measurement. The loss of the splitter along with cable loss is measured as -5.2 dB. The LO cable loss is 0.93 dB, RF cable loss is 0.94 dB and IF cable loss is 0.1 dB and splitter loss is 4.16 dB.\\

\subsection{Measurement setup for conversion loss}
Now insert the mixer in between MXG and VSA. We selected the RF frequency as 2.4GHz with power level of -25 dBm and LO signal of 10 dBm at 2.235 GHz, which provides the IF value of 165 MHz. we measured the IF value at 165 MHz, which results as -34.4 dBm provides the conversion loss of 9.3dB. The following table gives the conversion loss at various power-levels. 
\begin{table}[h!]
\begin{center}
 \begin{tabular}{|c|c|} 
 \hline
 RF Pin (dBm) & IF Out Power (dBm)  \\ 
 \hline
 -25.1 & -34.4 \\ 
 \hline
 -15.1 & -24.3  \\
 \hline
 -5.1 & -14.14  \\
 \hline
 4.9 & -4.7  \\
 \hline
 5.9 & -3.8  \\
 \hline
 6.9 & -3.15  \\
 \hline
 7.9 & -2.5  \\
 \hline
\end{tabular} \\
\end{center} 
\caption{IF Power vs RF Pin}
\end{table}\\

The following figure shows the conversion loss vs RF power. Here we plotted interpolated line from first RF power to show the conversion loss in increases as the RF power increases which is nothing but compression at higher RF powers. The ip1 dB is 7dBm. 
\begin{figure}[h!]
\centering
\includegraphics[width=8cm]{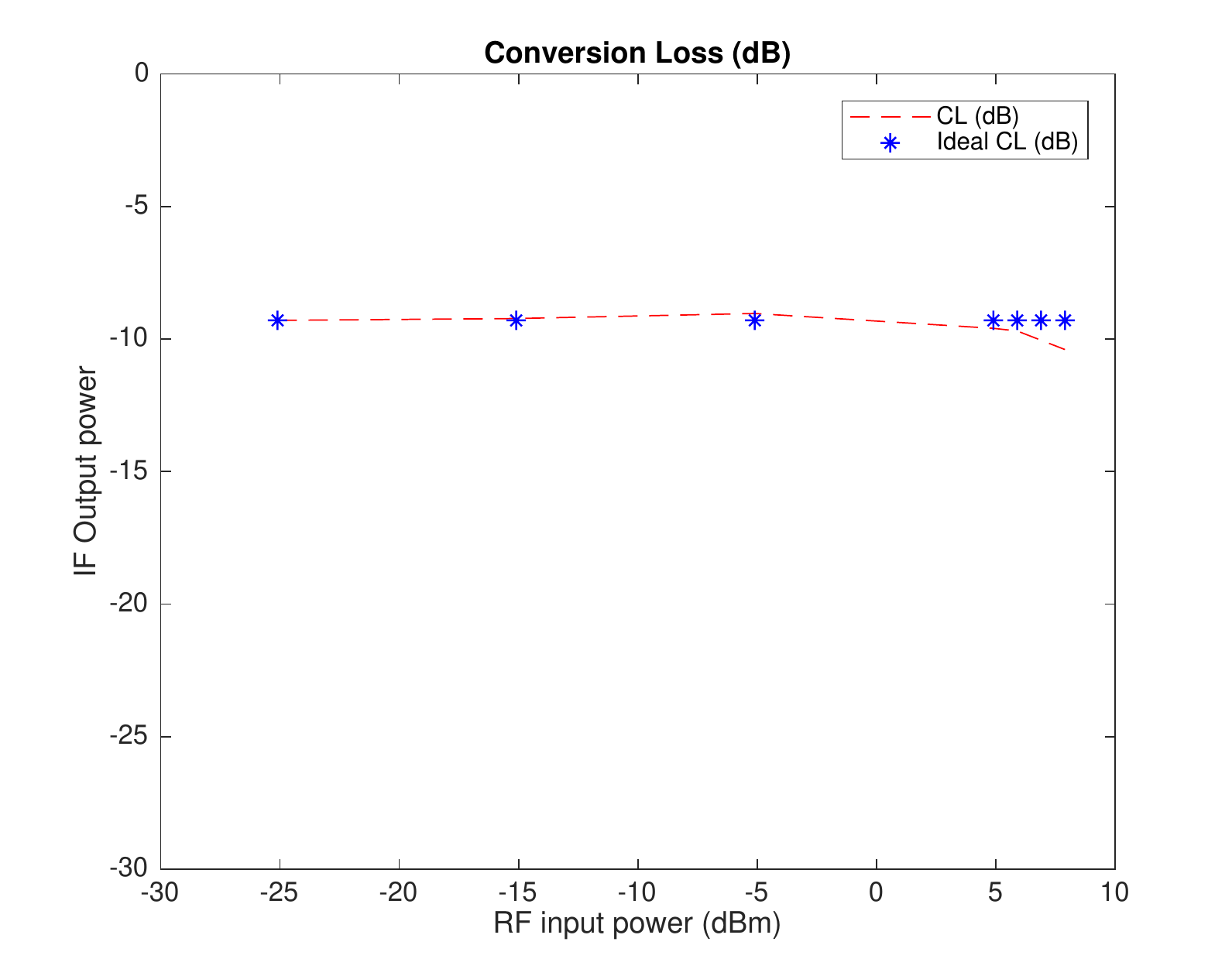}
\caption{Conversion loss vs RF input power (dB)}
\end{figure} \\

\subsection{Measurement setup for Isolation}
In this step, we measured the isolation across all ports. The RF to IF isolation is measured such that apply the tone at RF port and measured the tone at IF port with RF frequency. Example, apply a tone of -25.1 dBm at RF input @2.4 GHz and measure the IF @2.4 GHz (in this case, no LO signal is applied). The RF to IF isolation is measured as 17.6 dB. \\
\begin{table}[h!]
\begin{center}
 \begin{tabular}{|c|c|} 
 \hline
 RF Pin (dBm) & IF Out Power (dBm @2.4 GHz) \\ 
 \hline
 -25.1 & -47.2 \\ 
 \hline
 -15.1 & -32.72  \\
 \hline
 \end{tabular} \\
\end{center} 
\caption{RF vs IF isolation}
\end{table}\\
The LO to IF isolation is measured such that apply the tone at LO port and measured the tone at IF port with LO frequency. Example, apply a tone of 4.9 dBm at LO input @2.235 GHz and measure the IF @2.235 GHz (in this case, no RF signal is applied). The LO to IF isolation is measured as 20.3 dB.\\

In the same way, the LO to RF isolation is measured such that apply the tone at LO port and measured the tone at RF port with LO frequency. Example, apply a tone of 4.9 dBm at LO input @2.235 GHz and measure the RF port @2.235 GHz (in this case, IF port is terminated with matched load). The LO to RF isolation is measured as 6.4 dB. \\

\section{Non-linear measurements}
In this section we perform the non-linear measurements of the designed mixer using two-tone test. In this section, we discussed mainly about the third order inter-modulation power. The measurement setup is same as our-previous lab setup. \\
\subsection{Losses measured}
We measured the total loss with and without splitter. We got the loss due to each component. The splitter gave the loss of 4.1 dB, the RF cable has loss of 0.94 dB, LO cable has loss of 0.93 dB and IF cable has loss of 0.1 dB. These losses are same in the above mentioned section. \\ 
\subsection{IM3 vs RF power}
We used a two tone (2-tone) inputs at same power levels and measured the Pout at inter-mods using the PXA. For RF input power, we used 2.4 GHz for generator 1 and 2.45 GHz for generator 2. The inter-mod product is observed at 2.5 GHz. Initially we increased the input power in steps of 5 dBm, but later input power is increased in steps of 1 dBm.  To find the IM3 point, the Pout at f1 and IM3 at 2f2-f1 power levels should meet. We interpolated this data for more number of power levels. \\
The Third-order inter-modulations curve is derived based on few observations and interpolated throughout the Pin (dBm). 
The following table shows the IM3 powers vs RF input powers. 
\begin{table}[h!]
\begin{center}
 \begin{tabular}{|c|c|} 
 \hline
 Pin (dBm) & IM3 Power (dBm)  \\ 
 \hline
 -5.1 & -71.7 \\ 
 \hline
 -0.1 & -56.5  \\
 \hline
 4.9 & --42.2  \\
 \hline
 5.9 & -40.2  \\
 \hline
 6.9 & -38.4  \\
 \hline
 7.9 & -36.7  \\
 \hline
\end{tabular} \\
\end{center} 
\caption{IM3 Power vs Pin}
\end{table}\\

The following figure shows IIP3 of the mixer by interpolating the RF input data, RF output data and IM3 data. 
\begin{figure}[h!]
\includegraphics[width=9cm]{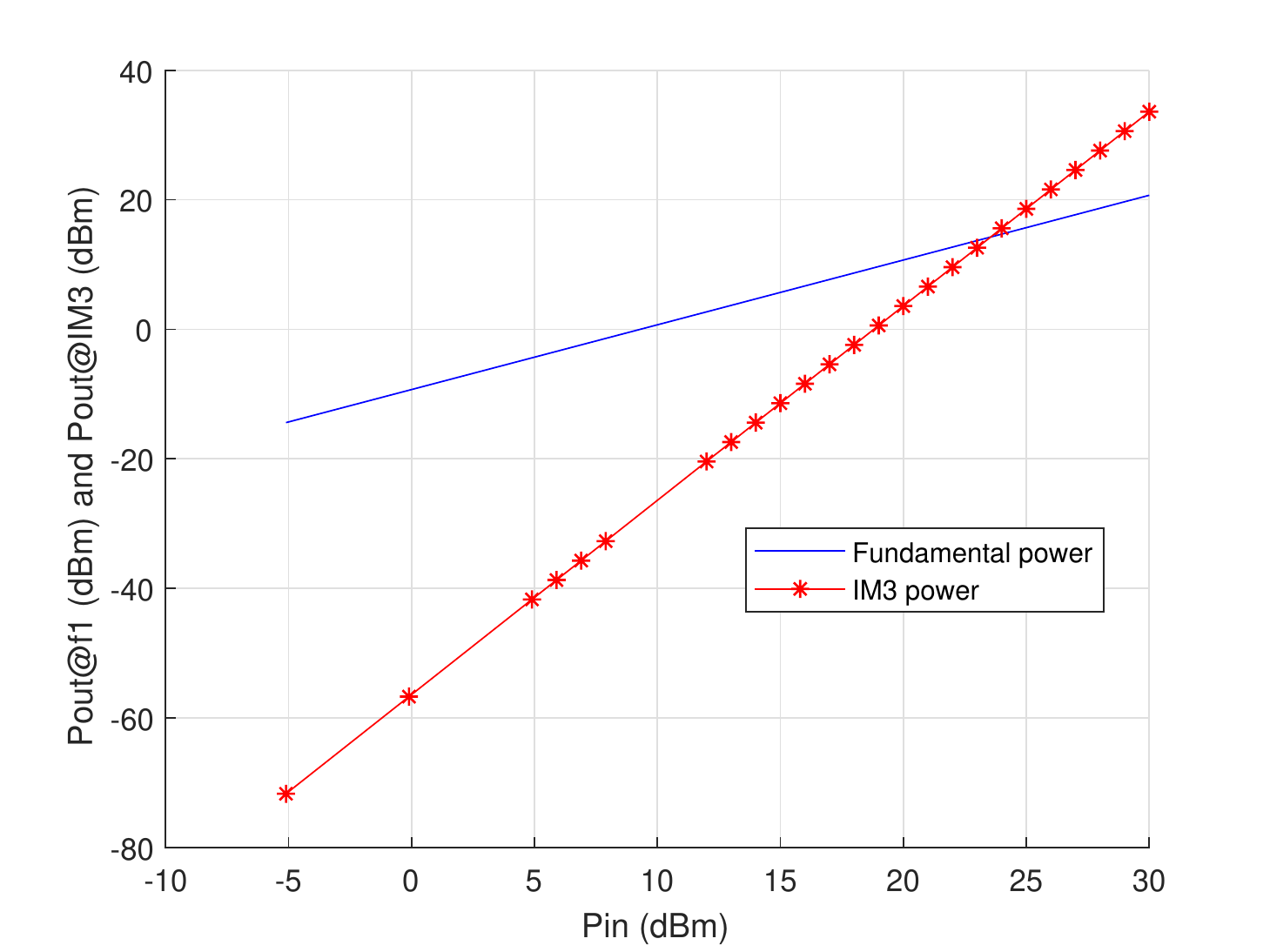}
\caption{Third-order inter-modulations vs Pin}
\end{figure}\\

We observed the OP1dB and IM3 power are intersecting at input power of 23.5 dBm as IIP3 value and 14.1 dBm as OIP3 values. \\

\section{Matching of the Mixer}
In this section, we found the match of the mixer circuit from all ports using ADS simulation. 
After the mixer designed, we did the S-parameters of the RF, LO and IF ports to find the refections from each of the port. The following figure shows the return loss of the RF port ($S_{11}$). \\ 
\begin{figure}[h!]
\includegraphics[width=8cm,height=6cm]{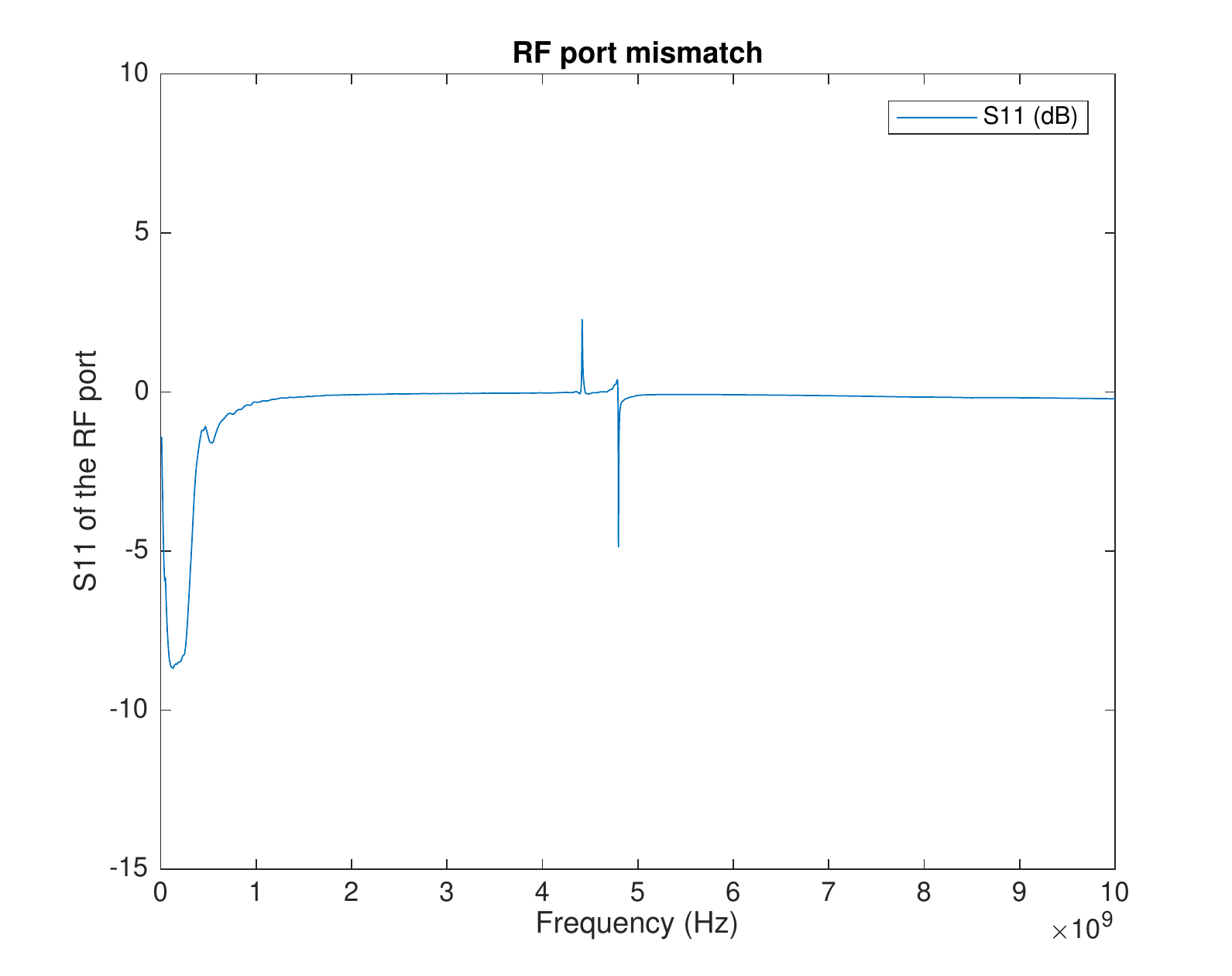}
\caption{Return loss of the RF port}
\end{figure} \\

The following figure shows the return loss of the LO port ($S_{22}$). \\
\begin{figure}[h!]
\includegraphics[width=8cm,,height=6cm]{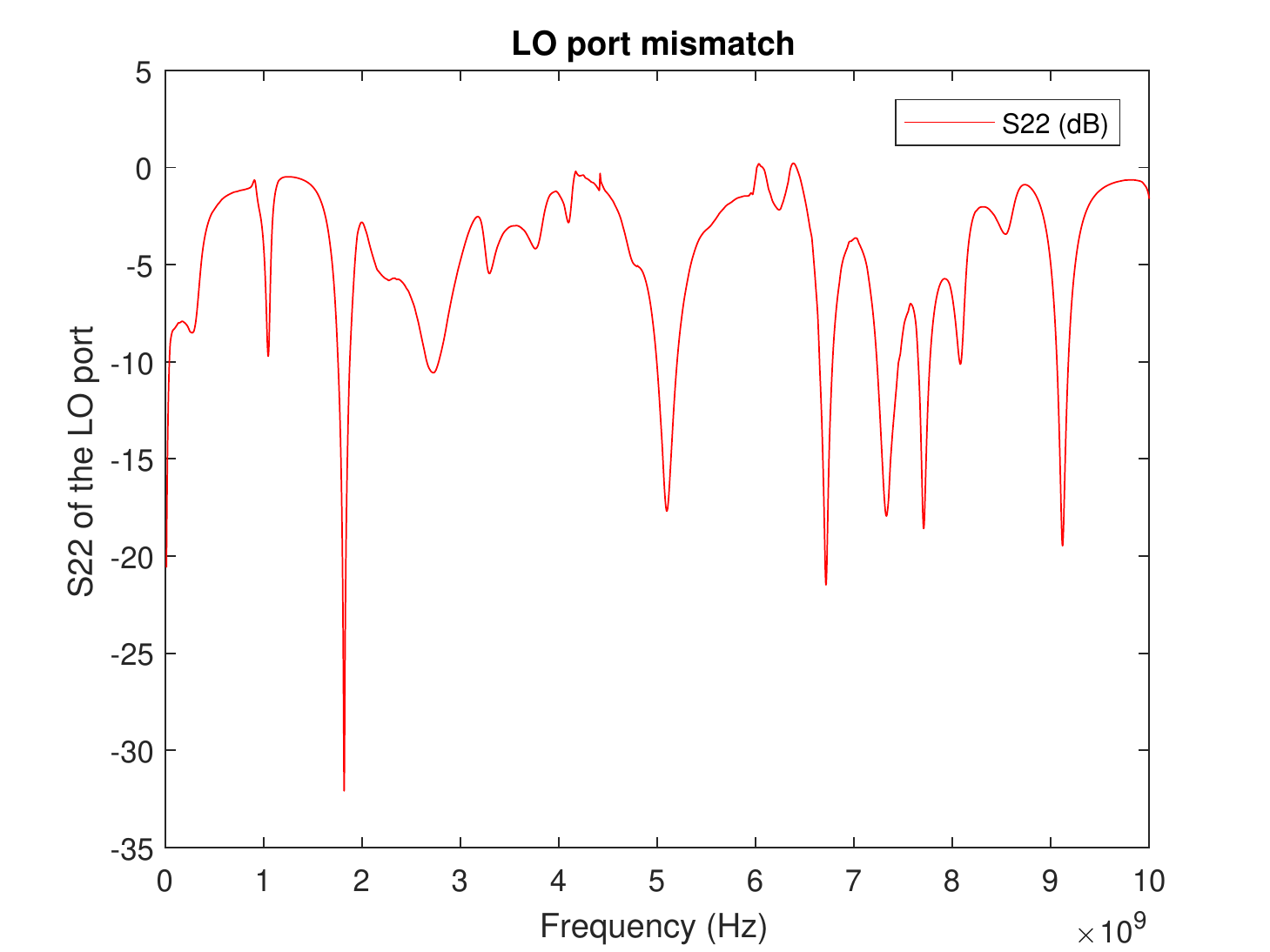}
\caption{Return loss of the LO port}
\end{figure} \\

The following figure shows the return loss of the IF port ($S_{33}$). \\
\begin{figure}[h!]
\includegraphics[width=8cm,,height=6cm]{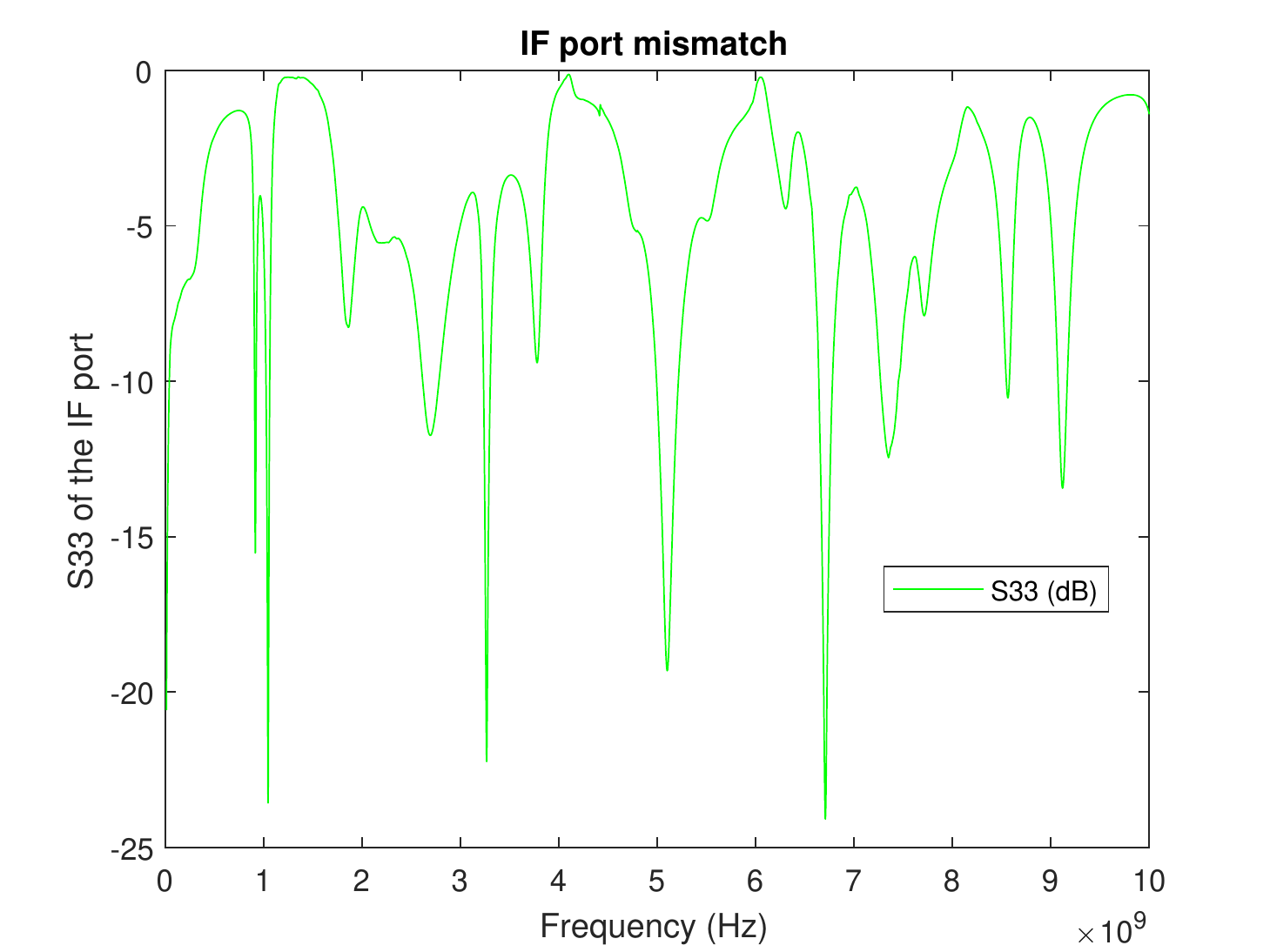}
\caption{Return loss of the IF port}
\end{figure}

\section{CONCLUSIONS}
We designed the single balanced mixer from scratch and understood each component in the design which effects the mixer performance. We fabricated the mixer using LPKF machine with Rogers 4350B substrate with Effective permittivity is 3.745. We performed the mixer linear and non-linear measurements. We understood how to measure the conversion loss of the mixer using the VSA. We verified the measured values are closely matching with the data sheet values. This gave us confidence to design and measure any active component in the next labs. Coming to conversion loss, we measured it which is 9.3 dB. For non-linear measurements, we used two-tone tests, same as with our previous labs. The compression point and IM3 points are measured and verified using data sheets, here the IIP3 is 17.9 dBm. 


%%%%%%%%%%%%%%%%%%%%%%%%%%%%%%%%%%%%%%%%%%%%%%%%%%%%%%%%%%%%%%%%%%%%%%%%%%%%%%%%

%%%%%%%%%%%%%%%%%%%%%%%%%%%%%%%%%%%%%%%%%%%%%%%%%%%%%%%%%%%%%%%%%%%%%%%%%%%%%%%%

%%%%%%%%%%%%%%%%%%%%%%%%%%%%%%%%%%%%%%%%%%%%%%%%%%%%%%%%%%%%%%%%%%%%%%%%%%%%%%%%
%\section*{APPENDIX}


%\section*{ACKNOWLEDGMENT}

%The preferred spelling of the word ÒacknowledgmentÓ in America is without an ÒeÓ after the ÒgÓ. Avoid the stilted expression, ÒOne of us (R. B. G.) thanks . . .Ó  Instead, try ÒR. B. G. thanksÓ. Put sponsor acknowledgments in the unnumbered footnote on the first page.

%%%%%%%%%%%%%%%%%%%%%%%%%%%%%%%%%%%%%%%%%%%%%%%%%%%%%%%%%%%%%%%%%%%%%%%%%%%%%%%%

%References are important to the reader; therefore, each citation must be complete and correct. If at all possible, references should be commonly available publications.

\end{document}